\pdfoutput=1
\documentclass[12pt]{article}

\usepackage{jheppub}
\usepackage{amsmath}
\usepackage{amssymb}
\usepackage{epsfig,amssymb,amsmath,psfrag,hyperref}
\usepackage[dvipsnames]{xcolor}
\usepackage{float} 
\usepackage{bbm} 
\usepackage{mathtools} 
\usepackage[enableskew]{youngtab} 
\usepackage{ytableau} 
\usepackage{slashed} 
\numberwithin{equation}{section}
\usepackage{caption}
\usepackage{booktabs}
\usepackage{mwe}
\usepackage{gensymb}
\usepackage{ulem}
\usepackage{cleveref}
\usepackage[toc,page]{appendix}
\usepackage{xcolor}

\usepackage{graphicx}

\usepackage{multirow, graphicx,amssymb,url,mathrsfs,amsmath}
\usepackage{eucal,wrapfig,boxedminipage,setspace,subfigure}
\usepackage{amsxtra,amstext,latexsym,dsfont,xcolor}

\newcommand{\ii}{\mathrm{i} }

               \def\D  {\Delta}

\def\IR{{\hbox{{\rm I}\kern-.2em\hbox{\rm R}}}}
\def\IB{{\hbox{{\rm I}\kern-.2em\hbox{\rm B}}}}
\def\IN{{\hbox{{\rm I}\kern-.2em\hbox{\rm N}}}}
\def\IC{\,\,{\hbox{{\rm I}\kern-.59em\hbox{\bf C}}}}
\def\IZ{{\hbox{{\rm Z}\kern-.4em\hbox{\rm Z}}}}
\def\IP{{\hbox{{\rm I}\kern-.2em\hbox{\rm P}}}}
\def\IH{{\hbox{{\rm I}\kern-.4em\hbox{\rm H}}}}
\def\ID{{\hbox{{\rm I}\kern-.2em\hbox{\rm D}}}}

\def\Tr{{\rm Tr}\,}

\def\det{{\rm det}}

\def\N{{\mathcal{N}}}

\newcommand{\beq}{\begin{equation}}
\newcommand{\eeq}{\end{equation}}
\newcommand{\bea}{\begin{eqnarray}}
\newcommand{\eea}{\end{eqnarray}}

\DeclareMathOperator{\STr}{STr}
\usepackage{cleveref}
\newcommand{\lrb}[1]{\left(#1 \right)}
\newcommand{\lrmb}[1]{\left[#1 \right]}
\newcommand{\lrbb}[1]{\left\{#1 \right\}}
\newcommand{\pd}{\partial}
\newcommand{\invrud}{\lrb{{1\over r_u^4}+{1\over r_d^4}}}
\newcommand{\pdr}{\pd_\rho}
\newcommand{\pdbr}[1]{\pdr\left(\rho^3\pdr #1\right)}
\newcommand{\lu}{L_u(\rho)}
\newcommand{\ld}{L_d(\rho)}
\newcommand{\lun}{L_u}
\newcommand{\ldn}{L_d}
\newcommand{\ludn}{L_{u/d}}

\newcommand{\id}{\mathbbm{1}}
\newcommand{\myslash} [1] {#1 \kern-.5em/}

\begin{document}

$\left. \right.$   \vspace{-18cm}

\title{$\left. \right.$ \vspace{17cm} \\ Holographic
Non-Abelian Flavour  Symmetry Breaking}
\author[a]{Johanna Erdmenger,}
\author[b]{Nick Evans,}
\author[a]{Yang Liu} 
\author[a]{and Werner Porod}

\affiliation[a]{Institute for Theoretical Physics and Astrophysics,
Julius-Maximilians-Universit{\"a}t W{\"u}rzburg,\\
97074 W{\"u}rzburg, Germany.}
\affiliation[b]{School of Physics \& Astronomy and STAG Research Centre, University of Southampton, Highfield,  Southampton  SO$17$ $1$BJ, UK.}
\emailAdd{erdmenger@physik.uni-wuerzburg.de}
\emailAdd{n.j.evans@soton.ac.uk}
\emailAdd{yang.liu@physik.uni-wuerzburg.de}
\emailAdd{porod@physik.uni-wuerzburg.de}

\abstract{We investigate a holographic model for both  spontaneous and explicit symmetry breaking of non-abelian flavour symmetries. This consists of a bottom-up model inspired by the top-down  D3/D7 probe brane model that incorporates the running anomalous dimensions of the fields.  We ensure that in the holographic bulk, the full non-abelian flavour symmetries for massless quarks are present. The quark masses are spontaneously generated field values in the bulk and there is a resultant bulk Higgs mechanism. We provide a numerical technique to find the mass eigenvalues for a system of coupled holographic fields. We test this approach using an analytic model of ${\cal N}=2$ supersymmetric matter. We apply this approach to two-flavour QCD with both $u-d$ quark mass splitting and multi-trace bulk action terms that are expected to break $U(N_f)_V$ to SU($N_f)_V \times U(1)_V$ away from large $N_c$. We also discuss three-flavour QCD with strange quark mass splitting and applications to more exotic symmetry breaking patterns of potential relevance for composite Higgs models.}

\maketitle

\setcounter{page}{1}\setcounter{footnote}{0}

\section{Introduction \label{intro}}

Generalizations of the AdS/CFT correspondence \cite{Witten:1998qj,Maldacena:1997re} have provided a new perspective on strongly coupled QCD-like gauge theories with confinement and chiral symmetry breaking. A number of approaches have been followed, both using top-down D-brane constructions \cite{Babington:2003vm, Kruczenski_2003, Sakai:2004cn, deTeramond:2005su, Karch:2002sh, Kruczenski:2003be,Kruczenski:2003uq} and bottom-up effective gravity actions \cite{Erlich:2005qh,DaRold:2005mxj,Arean:2012mq,Erdmenger:2020flu,Erdmenger:2014fxa,Erdmenger:2020lvq,Gursoy:2007cb,Gursoy:2007er,Jarvinen:2011qe,Jarvinen:2015ofa, Evans:2013vca,Alho:2013dka}. 
Gauge/gravity duality concepts were applied to QCD-like theories 
to address chiral symmetry breaking  for instance in \cite{Babington:2003vm,Kruczenski:2003uq,Sakai:2004cn}. Meson masses were calculated in the approach in \cite{Erlich:2005qh,DaRold:2005mxj,Erdmenger:2007cm,Arean:2012mq}, and baryon masses in\cite{deTeramond:2005su,Erdmenger:2020flu}.
These approaches provide sensible predictions for the meson spectrum and couplings, at least at the 15\% level, or even better
\cite{Erdmenger:2007cm,Bali:2007kt,Erdmenger:2014fxa,Erdmenger:2020flu,Erdmenger:2020lvq}. 
Moreover, the results compare favourably  to lattice studies  \cite{Erdmenger:2007cm,Bali:2007kt,Erdmenger:2014fxa,Erdmenger:2020flu,Erdmenger:2020lvq}. 

These 
 holographic techniques were extended to other non-abelian gauge theories \cite{Jarvinen:2011qe, Evans:2013vca, Alho:2013dka,Elander:2020nyd,Elander:2021bmt}. It is natural to also apply them to strongly coupled models of Beyond the Standard Model (BSM) physics. For example, holographic studies of  technicolour and Composite Higgs
 were performed in \cite{Contino:2003ve,Hong:2006si,Hirn:2006wg,Hirn:2006nt,Carone:2006wj,Agashe:2007mc,Haba:2008nz,Alho:2013dka,Belyaev:2019ybr,Elander:2022ebt,Elander:2023aow}. Recently, some of the authors of the present paper have used a bottom-up holographic approach that retains some essential features of the D3/D7 top-down probe brane model
\cite{Abt:2019tas,Erdmenger:2020lvq,Erdmenger:2020flu} to investigate the meson spectrum and
the top partner baryons for a large class of Composite Higgs  models presented in  \cite{Barnard:2013zea,Ferretti:2013kya,Ferretti:2014qta,Ferretti:2016upr}.

Given these successes, there has been only a small amount of studies on realizing non-abelian flavour symmetries with multiple quarks of different mass. In many cases where multiple quarks are included they are assumed to be degenerate - the computations reduce to those for a single quark flavour with the non-abelian symmetry simply allowing one to assert that the full U($N_f$) multiplets have the same mass and couplings. 
The full non-abelian structure is needed when considering flavour symmetry breaking via different quark masses or interactions. Some of the basic features of our bottom-up model were considered already in \cite{Shock:2006qy}. Also in the context of the Sakai-Sugimoto model \cite{Sakai:2004cn}, baryonic states with different quark masses were obtained, for instance in \cite{Hashimoto:2009st,Liu:2022urb}.

The aim of this present paper is to extend the holographic framework inspired by the D3/D7-brane construction beyond an axial $U(1)$ to include non-abelian flavour structures. In particular, we  stress the need for bulk gauge fields for the full non-abelian flavour symmetry of the massless theory - quark masses are values of supergravity fields in the bulk and so must be considered to spontaneously break these symmetries in the full bulk description. A common feature of these models is that one ends up with bulk theories with mixed fields - we show how to numerically extract the mass eigenvalues in this case and use an analytically solvable model of meson masses in an ${\cal N}=2$ theory to demonstrate it. We will then apply these ideas to QCD and begin to consider models with more elaborate global symmetry breaking patterns.

Amongst the top-down string constructions based on adding flavour through probe branes \cite{Karch:2002sh}, one possibility to obtain chiral symmetry breaking ($\chi$SB)  is to  embed D7-brane probes into supersymmetry breaking backgrounds. The model \cite{Babington:2003vm} describes an RG flow from a four-dimensional conformal field theory, $\N=4$ Super Yang-Mills theory broken to $\N=2$ through an additional supermultiplet in the fundamental representation of the gauge group, to a confining theory with $\chi$SB in the infrared. It is straightforward to dial the quark mass using the asymptotic boundary conditions on the probe brane embedding. In this way, explicit symmetry breaking effects are easily included. Here though adding additional flavours does not enhance the axial $U(1)$ to $U(N_f$) since the quarks all have a Yukawa interaction to a single adjoint scalar.  

In ref.~\cite{Erdmenger:2007vj}, a top-down
inspired D3/D7 brane model was presented with the aim to holographically describe mesons consisting of a heavy and a light quark using a non-abelian DBI action. Although this model still only has a U(1) axial symmetry, the discussion centred on the breaking of SU($N_f$) vector by the quark masses including a bulk Higgs mechanism.
We take this model as a starting point - we point out that the base quadratic kinetic terms do possess a full U($N_f)_L \times$ U($N_f)_R$ flavour symmetry but it is broken by the scalar potential to U($N_f)_V$. We therefore simplify the model to the quadratic order kinetic terms and construct models by adding different potentials that lead to different symmetry breaking patterns. Bulk gauge fields for the global symmetry of the massless model must be introduced.

Our first example is essentially to reconstruct the DBI model of the ${\cal N}=2$ supersymmetric theory of \cite{Erdmenger:2007vj}. We include the potential from the Dirac-Born-Infeld (DBI) case that breaks the global symmetry group to SU($N)_V\times U(1)_A$.  We use this case to explore the problem of mixed fields in the bulk - when there is quark mass splitting there is a choice of basis states. For example with two flavours, in the scalar sector, one can use the $\bar{u}u$ and $\bar{d}d$ basis (at large $N_c$ this is the physical mass eigenstate basis) or the basis $1/\sqrt{2}(\bar{u}u \pm \bar{d}d$). The mass eigenstates are independent of the basis, of course, but we are careful to learn how to compute if one chooses a basis with mixing. We find a numerical prescription to identify the mass eigenstates and verify it against the analytic solutions. We also discuss the Higgs mechanism in the bulk when the $u$ and $d$ masses split showing  the model reproduces the equations of motion for the mesonic states made of $\bar{u}d$ quarks found in \cite{Erdmenger:2007vj}.

Our main example in this paper is to apply our framework to large $N$ QCD: we introduce a potential that preserves SU(2)$_L \times$SU(2)$_R$, but generates a quark condensate that breaks the global group SU(2)$_L \times$~SU(2)$_R$ to SU(2)$_V$ in the massless limit. Quark masses are introduced through appropriate UV boundary conditions. Much of this is familiar from AdS/QCD models but it is important to flesh out this framework in the context of these models. 
We again exhibits a Higgs mechanism in the bulk for all symmetry breaking  patterns of the model which leads, for example, to splitting of the neutral and charged pion masses. 

Our main original motivation for this work was to explore the breaking of U(2)$_V$ symmetry to SU(2)$_V \times$ U(1)$_V$ by the inclusion of double-trace terms in the potential of the fields in the bulk. These terms are expected away from the $N_c \rightarrow \infty$ limit. Double tarce terms in the bulk potenital can realise this splitting in the scalar sector. We show they split the $\sigma= 1/\sqrt{2}(\bar{u}u + \bar{d}d)$ isospin singlet and the $\xi^i=1/\sqrt{2}(\bar{u}u - \bar{d}d), \bar{u}d, \bar{d}u$ multiplet of isospin triplet scalars. If both the double trace term and quark masses are present then the bulk has field mixing and the mass eigenstates are not simple to spot a priori - we use our numerical methods to extract the mass eigenstates in this case.

As a further example we discuss also QCD with three flavours.  Here we focus mainly on the fact that $m_s\gg m_u,m_d$ and set for simplicity set $m_u=m_d$. We obtain masses for the pseude-Nambu-Goldstone Bosons (pNGBs), the scalar bound states, the vector mesons and axial vector mesons. Our results agree in general quite well with their measured values. 

We conclude this paper with an outline of how to treat other more exotic flavour group breaking patterns relevant for model building beyond the Standard Model. We leave studying particular cases for future work, having established the framework and computational tools here.

This paper is organized as follows: In Section \ref{DBI_JE} we review the non-abelian top-down model as reported in \cite{Erdmenger:2007vj}. In Section \ref{sec:n2 model} we recreate that model in the bottom-up framework and explore the calculational tools necessary to extract the meson spectrum. We then extend it to QCD in Section \ref{sec:our model}, discussing the scenarios with degenerate and non-degenerate quarks in the two- and three-flavour theory. In Section \ref{Other_section} we discuss the incorporation of more elaborate symmetry breaking patterns. We draw our conclusions in Section \ref{conclusion} and discuss potential future work.

\section{Non-Abelian DBI Action for the D3/probe D7 System (Summary) \label{DBI_JE}}

We collect here the key elements of
the non-abelian Dirac-Born-Infeld (DBI) description of ref.~\cite{Erdmenger:2007vj} for the convenience of the reader. That model describes $N_f$ flavours of ${\cal N} =2$ supersymmetric quark hypermultiplets interacting with the glue sector of an ${\cal N}=4$ gauge theory. The model, by virtue of the Yukawa terms between the quarks and the adjoint scalar fields, has only a U($N_f$) $\times$ U(1)$_A$ symmetry. This holographic dual description serves as the ingredients for the model presented in Sections \ref{sec:n2 model}-\ref{Other_section}.

 The ${\cal N}=4$ SYM theory has a dual described by 
the AdS$_5 \times S^5$ metric which is conveniently written as
\begin{equation}
ds^2 = r^2 d x_{3+1}^2 + {1 \over r^2} \left[ d\rho^2 + \rho^2 d \Omega_3^2 + dL^2 + L^2 d\theta^2 \right]  \, ,
\end{equation}
using coordinates appropriate for the embedding of a D7-brane probe. 

The starting point to describe the quark dynamics is the non-Abelian Dirac-Born-Infeld action proposed in ref.~\cite{Myers:1999ps},
\begin{align}
S_{N_f} = - \tau_p \int d^{p+1} \xi  e^{-\phi} \text{ STr} \left(
\sqrt{-\det ( P[G_{rs} + G_{ra} (Q^{-1} -\delta)^{ab} G_{sb}] 
+ T^{-1} F_{rs})}
\sqrt{\det Q^a{}_b} \right)  
\label{eq:non-abelian_DBI}
 \,.
\end{align}
It describes the dynamics of $N_f$
D$p$-branes in a background with metric $G_{mn}$. $\phi$ is the dilaton, $F_{rs}$ the  world-volume field strength tensor and $T^{-1} =2\pi\alpha '$. 
 It is important to note that the metric elements $G_{ab}(r^2)$ have
a matrix structure, e.g. for the case of diagonal real masses we shall use below
\begin{equation}
\label{eq:matrixr}
    G_{\rho\rho} = \frac{1}{r^2}  \quad
    \text{with} \quad
    r^2 = \begin{pmatrix}
    r^2_{u} & 0\\ 0 & r^2_{d}  
    \end{pmatrix}\,.
\end{equation}

The matrix $Q$ is defined by
\begin{align}
  Q^a{}_b = \delta^a{}_b + \ii \,T \,[X^a,X^c] G_{cb} \,   \end{align}
where $X^a$ are the coordinates transverse 
to the stack of D$p$-branes. These take values in an  $U(N_f)$ algebra.
 All the fields $X^a, A^a$ and the metric elements $G_{ab}$ transform in the adjoint of U($N_f$) transforming as $F \rightarrow U^\dagger F U$ where $F$ is a generic field and $U$ an element of the vector U($N_f$) global symmetry.

The symbol ${\rm STr}$ denotes the
symmetrized trace 
\begin{align}
\text{STr}(A_1 ...A_n) \equiv \frac{1}{ n!} \sum\text{Tr}\left(A_1...A_n  +\text{ all permutations}\right)     
\label{eq:str}
\end{align}  
  and is needed to avoid the
ordering ambiguity of the expansion of the DBI action \cite{Tseytlin_1997}. We note two technical details for completeness: (i) commutators of Lie-algebra valued objects are considered as one matrix $A_i$ in eq.~\eqref{eq:str} \cite{Tseytlin_1997,Myers:1999ps,Denef:2000rj}. (ii) one has first to sum over the space-time indices before performing the symmetrized trace over the Lie-algebra valued objects \cite{Denef:2000rj}. 
 We will discuss this prescription in more detail in our bottom-up models below.

In our convention, $r,s= 0,1,...,p$ and $a,b= p+1, ..., 9$ label the
world-volume directions and the directions transverse to the D$p$-branes,
respectively; $m,n=0,1, \cdots, 9$ are the 10d spacetime indices. In the following we take $p=7$.
$P[A_{rs}]$~denotes the pull-back of a 10d tensor $A_{mn}$ to the
world-volume of the branes which is given by
the covariant derivative in case of
the non-Abelian DBI,
\begin{align}
 D_r X^a=\partial_{r}X_a + \ii [A_r, X^a]   
\end{align}
with 
non-Abelian world-volume gauge field $A_r$.

As in \cite{Erdmenger:2007vj}, we consider a diagonal brane embedding. The action is then expanded in powers of $[X^a, X^b]$,
leading to 
\begin{align}
 S_{N_f}= 
  \tau_p   \int d^{8} \xi \
 e^{-\phi}\text{STr}  &\bigg\{  
  \sqrt {-\det 
({G}_{rs}  + G_{ab}D_r X^a D_s X^b
 +   T^{-1} F_{rs})} \nonumber \\
 &\cdot  \left(1-\frac{1}{4}\left(T\,G_{ac} [X^c,X^b]\right)^2\right) \ \bigg\} + {\cal O}\left( X^4\right)\ .
 \label{eq:non-abelian_DBI_expanded1}
\end{align}
According to~\cite{Erdmenger:2007vj},the 
diagonal ansatz for the embeddings leads
to a significant simplication in the
determination of the embedding functions
for the $N_f$ probe D7 branes in different gravity backgrounds. The diagonal ansatz is
given by
\begin{align}
  X^a= \text{diag}(L_1^a, \cdots, L_{N_f}^a) \quad\quad (a=8,9)
\end{align}
and leads to
\begin{align}
 S_{N_f} = \tau_7   \int d^{8} \xi \  e^{-\Phi}\sum_{i=1}^{N_f}  
  \sqrt {-\det ({G}_{rs}+ G_{ab}\partial_r L_i^a \partial_s L_i^b )}    \,.
\end{align}
with the metric factors of the form in eq.~ (\ref{eq:matrixr}).
Thus, we obtain $N_f$ decoupled equations of motion for the $L_i^a$. Their
explicit form will of course depend on the metric $G$.
The asymptotic value of the
$L_i^a$ in the ultraviolet limit is given by
the corresponding quark mass for each flavour. 
The details for the resulting action were worked out in ref.~\cite{Erdmenger:2007vj} for the case of an $U(2)$ group. 

For the fluctuations perpendicular to the D7-branes, the ansatz of  \cite{Erdmenger:2007vj} is
\begin{align}
X^8 &= \tau_i \phi^8_i  \equiv \phi^8  \\
X^9 &= \begin{pmatrix}
      L_1 & 0\\
      0 & L_2\\
    \end{pmatrix} + \tau_i \phi^9_i    \, ,
\end{align}
where
  \begin{equation}
    \label{eq:U(2)_generator_basis}
    \tau^0=\frac{1}{2}
    \begin{pmatrix}
      1 & 0 \\
      0 & 1
    \end{pmatrix},\,
    \tau^1=\frac{1}{2}
    \begin{pmatrix}
      0 & 1 \\
      1 & 0
    \end{pmatrix},\,
    \tau^2=\frac{1}{2}
    \begin{pmatrix}
      0 & -i \\
      i & 0
    \end{pmatrix},\,
    \tau^3=\frac{1}{2}
    \begin{pmatrix}
      1 & 0 \\
      0 & -1
    \end{pmatrix}.
  \end{equation}
Expanding the integrand of the action
in eq.~\eqref{eq:non-abelian_DBI_expanded1}
to second order in the fields we obtain
\begin{align}
 {\cal L} = \rho^3 \STr\biggl( e^{\phi}\biggl[1&+ {1 \over 2} G_{\rho \rho} D^r X^{9\dagger} D_r X^9 +{1 \over 2} G_{\rho \rho}  \partial^r \phi^8 \partial_r \phi^8  + {1 \over 4 g^2}  F^{ab} F_{ab}\nonumber \\
 &+{1\over 8} G_{\rho\rho}^2(L_1-L_2)^2[(\phi^8_1)^2+(\phi^8_2)^2] \biggr]\biggr). 
\end{align}
The second line corresponds to the the second one in eq.~\eqref{eq:non-abelian_DBI_expanded1}, yielding a mass term for the scalar fluctuations in the 8-direction. The commutator structure
in eq.~\eqref{eq:non-abelian_DBI_expanded1} implies that a corresponding term is absent for $X^9$.

\section{A Bottom-Up Non-Abelian Model for the ${\cal N}=2$ Theory \label{sec:n2 model}}

 Our goal is to work towards describing an effective holographic description of any dynamical symmetry breaking pattern. To begin to establish the ground rules, we will start by creating a bottom-up holographic description for the ${\cal N}=2$ supersymmetric field theory described in the previous section, to demonstrate that our holographic approach captures the key elements of the dynamics. 

Let us begin by writing down a kinetic term for the field $X$ that determines the vacuum - we simply keep the quadratic order term from the DBI action

\begin{equation} \label{linDBI}
{\cal L}_{D7} = \rho^3 \STr \left[ {1 \over 2} G_{\rho \rho} G^{ab} \partial_a X^\dagger \partial_b X   \right], 
\end{equation}
where we have dropped a cosmological constant term and terms beyond quadratic order. It is helpful for concreteness to write $X$ an N$_f \times$ N$_f$ matrix as
\begin{equation}
\label{eq:general_param}
X = \left(L+\ \sum_k \xi_k(x) T_k \right)
 e^{i(\pi_a(x) T_a)} \,,
\end{equation}
here $L$ is a real diagonal matrix that encodes the vacuum values of $X$. The $T^a$ are the generators of U(N$_f$). The $ \xi^k$ and $\pi^a$ are then the 2N$_f$ components (they will become the fluctuations about the vacuum configuration $L$).

$G_{ab}$ is the AdS metric but written as a flavour matrix (in the brane language, pulled back onto the worldvolume of two separated D7s) and depends on the matrix $r^2=\rho^2 \id +L^2$. 

These kinetic terms have a full chiral flavour symmetry where
\begin{equation} 
X \rightarrow U_L^\dagger X U_R, \hspace{1cm}   G \rightarrow U_L^\dagger G U_R
\end{equation}
$U_L, U_R$ being the group actions of the chiral symmetries. The STr in the action implies averaging over all terms compatible with this symmetry. In particular we can form the two metric components as $G_{\rho \rho}^\dagger G^{ab}$ which transforms as ``$U_R U_R^\dagger$" and can be inserted in the trace at the beginning. Equally we can write $G^{ab \dagger} G_{\rho \rho}$ which transforms as ``$U_L U_L^\dagger$" and can inserted between $X^\dagger$ and $X$. We average over these possibilities.

Here we will restrict ourselves to considering cases where $L$ is real and diagonal, but not proportional to the identity. That is, the up and down quark masses will be unequal but both simultaneously real. The allowed vacuum configurations with $X$ non-zero are then given by two separated equations of motion with $X=$diag$(L_u, L_d)$ 
\begin{equation} \label{pso}
\partial_\rho [ \rho^3 \partial_\rho L_{u/d}] = 0, \hspace{1cm} L_{u/d} = m_{u/d} + { c_{u/d} \over \rho^2} 
\end{equation}
where in the solution shown $m_{u/d}$ is identified as being proportional to the two quark masses and $c_{u/d}$ to the quark condensates. 

The equations of motion follow from allowing $L_{u/d} \rightarrow L_{u/d} + \delta L_{u/d}$ and integrating by parts as usual. There are then also boundary terms from the variation of the action
\begin{equation} \label{bound}
S_b = \int d^4x {\partial {\cal L} \over \partial L} \delta {\cal L}_{u/d} = \int d^4x \rho^3 \delta L _{u/d}\partial_\rho  L_{u/d} 
\end{equation}
which are set to zero in the UV by fixing $L_{u/d}$ and in the IR by $L_{u/d}'(0)=0$. In the string picture, the IR condition is a regularity condition on the D7-brane embedding. One can also though view this IR condition as the result of imposing the surface potential $V= mc$ (ie eq.~(\ref{bound})  evaluated on the solution in eq.~(\ref{pso}) ) which enforces $c=0$ for any non-zero $m$ and also at $m=0$ through the limit of taking $m$ to zero.

The $X$ kinetic term has more symmetry than the ${\cal N}=2$ theory. To reduce the symmetry we include a suitable potential term to mimic the theories` moduli space
\begin{equation} \label{n2pot}
V = {1 \over 2}Tr [ X^\dagger,X]^2 = Tr(X^\dagger XX^\dagger X) - Tr(X^\dagger X^\dagger XX) 
\end{equation}
The second term explicitly breaks U($N_f)_L \times$ U($N_f)_R\rightarrow$ U($N_f$)$_V \times$U(1)$_A$. 

Next holography requires us to include a gauge field for the vector global symmetry (in the DBI picture this is the D7 worldvolume gauge field)  so our full kinetic terms become

\begin{equation} 
{\cal L} = \rho^3 STr \left[ {1 \over 2} G_{\rho \rho} G^{ab} D_a X^\dagger D_b X  + {1 \over 4 g^2} G^{ab} G^{cd} F_{ac} F_{bd} \right] 
\end{equation}
where
\begin{equation} \label{cd}
D_a X = \partial_a X + i [T^m, T^n] V_a^m X^n 
    \end{equation}

$F$ is again a flavour matrix transforming as $F\rightarrow V^\dagger F V$. The commutator coupling to $X$ reflects the vector nature of the symmetry - again we stress that the commutator must be performed before the STr. 

In the DBI picture the U(1)$_A$ symmetry is a remnant of the ${\cal N}=4$ gauge theories SU(4)$_R$ symmetry group. There is a gauge field in the AdS bulk that is dual to this symmetry. In the probe limit one normally neglects interactions with it. This is presumably the unique way to introduce the axial field and preserve ${\cal N}=4$ supersymmetry. We will therefore here neglect this field also so our effective theory mimics the DBI case.

\subsection{Example 1 - Equal, Real Masses}

The theory with  real and equal diagonal mass entries 
which corresponds to the background solution $X=m \id_{N_f \times N_f}$ is a very simple example.  We can write the real fluctuations simply as
\begin{equation}
X _{ij}= m \delta_{ij} + s_{ij} + i \tilde{s}_{ij}
\end{equation}
where $\delta_{ij}$ is the Kronecker delta. For this case the metric factors (setting any fluctuations to zero) are also proportional to the unit vector and for example
\begin{equation}
G^{\rho \rho} = G_{xx} = (\rho^2 + m^2 ) \id_{N_f \times N_f}
\end{equation}
Further the potential in eq.~(\ref{n2pot}) vanishes at quadratic order in the fluctuations (replace any two X in the first term with $m \id_{N_f \times N_f}$ and an equivalent second term cancels it). Finally the commutator with the vacuum $X$ in eq.~(\ref{cd}) vanishes and at quadratic order the vector and scalar fluctuations don't mix.

The upshot of all this is that in the scalar sector we obtain $2 N_f^2$ copies of the abelian equation
\begin{equation} \label{sc}
\partial_\rho[ \rho^3 \partial_\rho S] + {M^2 \over (\rho^2 + m^2)^2} S = 0
\end{equation}
and equally for the $N_f^2$ vector meson $V(\rho) e^{-i k.x}, M^2=-k^2$
\begin{equation} \label{ve}
\partial_\rho[ \rho^3 \partial_\rho V] + {M^2 \over (\rho^2 + m^2)^2} V = 0
\end{equation} These states are all degenerate and match those produced by the full DBI action.

\subsection{Example 2 - $N_f=2$ Split, Real Masses}

As our second example let's consider the two flavour case (we call them $u,d$) with the real, non-degenerate mass matrix diag($m_u,m_d)$. This contains many of the key ingredients of the non-abelian models. In the holographic model this corresponds to the diagonal vacuum solution $X_{\bar{u}u}=m_u$ and $X_{\bar dd} = m_d$ - both the kinetic and potential terms in the Lagrangian vanish on this solution for the vacuum. The solution  just corresponds to two D3 branes separated on the same axis and there is no quark condensate in the system. 

The meson mass spectrum can be split into two pieces - those associated with diagonal flavour matrices and those associated with off-diagonal matrices. 

\subsubsection{Diagonal States} 
\label{sec:diagonal_states}

In both the scalar and vector meson sectors the mass eigenstates for the diagonal fluctuations, on the introduction of mass splitting, immediately switch from being the isospin singlet and triplet elements to the simple $\bar{u}u$ and $\bar{d}d$ states. 
The $\sigma_u,\sigma_d$ basis is straightforwardly decribed. For example, the holographic action for the scalars is
\begin{equation}
S_{5d} = \int d^4x d \rho~ \rho^3 \left[ (\partial_\rho \sigma_u)^2  + {1 \over (\rho^2 + m_u^2)^2} (\partial_\mu \sigma_u)^2 + (\partial_\rho \sigma_d)^2  + {1 \over (\rho^2 + m_d^2)^2} (\partial_\mu \sigma_d)^2 \right].
\end{equation}
We solve the two equations assuming $\sigma_i(\rho,x)=\sigma_i(\rho)e^{iq.x},~ q^2=-M_i^2$
\begin{equation} 
\partial_\rho (\rho^3 \partial_\rho \sigma_{u/d} ) + \rho^3 {M_{u/d}^2 \over(\rho^2 + m_{u/d}^2)^2} \sigma_{u/d} = 0 
\label{eq:example}
\end{equation}
requiring $\sigma_{u/d} \rightarrow 0$ in the UV to fix $M^2_{u/d}$. 
We then substitute back into the action eg. for $\sigma_u(\rho,x)=\sigma_u(\rho) \sigma_u(x)$
\begin{equation} 
\begin{aligned}
S_{5d} =& \int d^4x d \rho  \left[ - \partial_\rho (\rho^3 \partial_\rho \sigma_u) \sigma_u + {\rho^3 (\partial_\mu \sigma_u)^2\over (\rho^2 + m_u^2)^2}  \right]\\ 
= &\int   d \rho~  {\rho^3 \sigma_u(\rho)^2 \over (\rho^2 + m_u^2)^2} \int d^4x ( (\partial_\mu \sigma_u(x))^2 + M_u^2 \sigma_u(x)^2)\end{aligned} \end{equation}
to normalize we then set 
\begin{equation}
\int d^4x d \rho~  {\rho^3 \sigma_{u/d}(\rho)^2 \over (\rho^2 + m_{u/d}^2)^2} = 1. \end{equation}

This has been just two copies of the Abelian case. The reason we stress this structure is that we now wish to show how one could have arrived in this basis if one had begun in a different basis where the states mix. For example,  the Lagrangian that emerges in the alternative basis $(\sigma,\tau)=1/\sqrt{2} (\sigma_u \pm\sigma_d)$ is 
\begin{equation} \begin{array}{ccc}
S_{5d} &=& \int d^4x d \rho~\left[\rho^3 (\partial_\rho \sigma)^2   +{\rho^3 \over 2} \left[ {1 \over (\rho^2 + m_u^2)^2} + {1 \over (\rho^2 +m_d^2)^2}  \right] (\partial_x \sigma)^2 \right.\\ &&\\
&& \left.  + \rho^3 (\partial_\rho \tau)^2  +{\rho^3 \over 2} \left[ {1 \over (\rho^2 + m_u^2)^2} + {1 \over (\rho^2 +m_d^2)^2}  \right] (\partial_x \tau)^2  \right. \\
&&\\
&& \left. +{\rho^3} \left[ {1 \over (\rho^2 + m_u^2)^2} - {1 \over (\rho^2 +m_d^2)^2}  \right] (\partial_x \sigma) (\partial_x \tau) \right].
\end{array}
\end{equation}

In reference \cite{Kaminski:2009dh}  a prescription to numerically find the mass eigenstates is provided that we take over to this case (there quasi-normal modes were considered). We should seek fluctuations of both $\sigma$ and $\tau$ that coherently have the form $e^{ikx}, k^2=-M^2$. 
This leads to the equation of motion 
\begin{equation} 
\partial_\rho (\rho^3 \partial_\rho \sigma) + {\rho^3 \over 2}\left[ {1 \over (\rho^2 + m_u^2)^2} + {1 \over (\rho^2 +m_d^2)^2}  \right] M^2 \sigma +{\rho^3 \over 2} \left[ {1 \over (\rho^2 + m_u^2)^2} - {1 \over (\rho^2 +m_d^2)^2}  \right] M^2 \tau = 0
\label{eq:set_nick1}
\end{equation}
\begin{equation} 
\partial_\rho (\rho^3 \partial_\rho \tau) + {\rho^3 \over 2}\left[ {1 \over (\rho^2 + m_u^2)^2} + {1 \over (\rho^2 +m_d^2)^2}  \right] M^2 \tau +{\rho^3 \over 2} \left[ {1 \over (\rho^2 + m_u^2)^2} - {1 \over (\rho^2 +m_d^2)^2}  \right] M^2\sigma = 0.
\label{eq:set_nick2}
\end{equation}

\begin{figure}
    \includegraphics[width=0.9\textwidth]{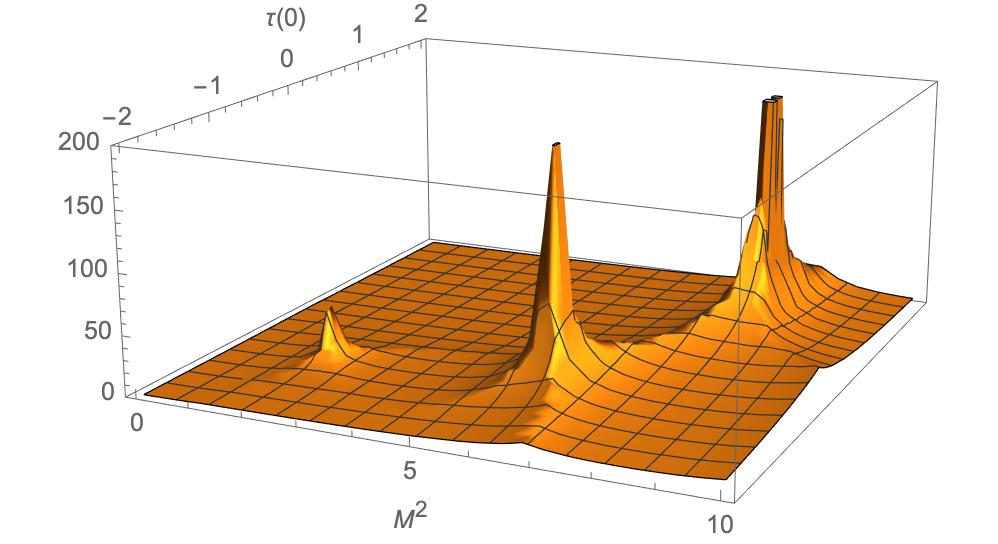}
      \caption{Here we show a numerical method to the solution of the mixed equations eq.~(\ref{eq:set_nick1}) and eq.~(\ref{eq:set_nick2}). The diagonal basis is known and the solutions in each of the $u$ and $d$ sectors are $M^2=4 (n+1)(n+2)m_q$. Here with fixed $\sigma(0)=1, \sigma'(0)=0, \tau'(0)=0$ we vary the mass squared and the value of $\tau(0)$. We plot the quantity $1/(|\sigma(\infty)| + |\tau(\infty)|$ which diverges when both the fields vanish asymptotically. In this case we have set one quark mass to unity and the other 0.5 - there are clear peaks at $\tau(0)=-1$ and $M^2=2,6$ which are the zeroth and first excited states of the $d$ quark state. The peak at $\tau(0)=1$ and $M^2=8$ is the ground state of the $u$ quark.}
    \label{fig:poles}
\end{figure}

To solve these numerically we start in the IR with boundary conditions $\sigma (0)=1$, $\sigma'(0)=0$ and $\tau'(0)=0$. Now we have two parameters $M^2$ and $\tau(0)$ that we can vary to seek solutions where both $\sigma$ and $\tau$ vanish in the UV. An example numerical method is shown in \cref{fig:poles}. 

In this case we can find these solutions analytically. They are $\sigma=\tau=\sigma_u$ with $M^2=M_u^2$ which returns our eq.~(2.23). Equally one can set $\sigma=-\tau=\sigma_d$ with $M^2=M_d^2$.

Now when we substitute back into the action we write eg. for the first case $\sigma=\sigma(\rho) \sigma_u(x)$ and $\tau=\tau(\rho) \sigma_u(x)$ and we find
\begin{equation} \begin{array}{ccc}
S_{5d} &=& \int d^4x d \rho~ {\rho^3 \over 2}\left[ {(\sigma+\tau)^2 \over (\rho^2 + m_u^2)^2} + {(\sigma-\tau)^2 \over (\rho^2 +m_d^2)^2}  \right]  \left[ (\partial_x \sigma_u)^2 +M_u\ \sigma_u^2 \right] \end{array}
\end{equation}
Note the mixed kinetic term is now no longer present because the two fields share the same x dependence and so that cross term just merges to form the diagonal kinetic term. Note that since $\sigma=\tau$ the $m_d$ dependence vanishes here.

\noindent\fbox{%
    \parbox{\textwidth}{%
        Summary of numerical method for mixed equations of motion: The general case is  if one has $n$ fields, $\sigma_i$, with mixed equations of motion. One should seek fluctuations of all $n$ fields that all have the form $e^{ikx}, k^2=-M^2$. To solve the equations numerically one shoots from the IR with boundary conditions $\sigma_1 (0)=1$, and $\sigma_i'(0)=0$ for all $i=1..n$. Now we have $n$ parameters $M^2$ and $\sigma_n(0)$ $i=2... n$ that we can vary to seek solutions where all $\sigma_i$ vanish in the UV. 
    }%
}

\subsubsection{Off-Diagonal States}

Meson states made from both $u$ and $d$ quarks follow a new story that was the focus of \cite{Erdmenger:2007vj}. Our model here contains all the key elements still.

Firstly consider the real off diagonal fluctuations of $X$ - since here $X$ is real, the potential in eq.~(\ref{n2pot}) vanishes. When the masses split in the vacuum solution for $X$, the commutator in eq.~(\ref{cd}) is non-zero with the off diagonal generators of SU(2)$_V$. The mass splitting breaks the U(2)$_V \rightarrow$ U(1)$^2$ and in the bulk there is a Higgs mechanism. 

To see this in detail, consider the particular case where we look at fluctuations associated with the $T^1$ generator and the $A^2$ component of the gauge field in flavour space,
\begin{equation} X = \left( \begin{array}{cc} m_u & \delta X_1 \\ \delta X _1&m_d \end{array} \right) \hspace{1cm} {\rm and} \hspace{1cm} A_2 = \left( \begin{array}{cc} 0 & -i  \\ i &0 \end{array} \right) \end{equation}
for which the covariant derivative in eq.~(\ref{cd}) becomes at linear order
\begin{equation} D_a X = (\partial_a \delta X _1 + (m_u-m_d) A_{2a})  \left( \begin{array}{cc} 0 & 1 \\ 1&0 \end{array} \right)  \, , \end{equation}
with $a$ the five-dimensional Lorentz index running over $(x^\mu, \rho)$.
In this case where the quark masses $m_u, m_d$ are constants, we can just make the gauge transformation
\begin{equation} A_{2a} \rightarrow \tilde{A}_{2a}  = A_{2a} - \partial_a {\delta X _1 \over (m_u-m_d) } \end{equation}
to  ``eat'' the $\delta X_1$ field.
On computing the kinetic term $|D_aX|^2$, we obtain a mass term for the gauge field $\tilde{A}_{2a}$ that is proportional to $(m_u-m_d)^2$. The gauge field kinetic term is of course, by construction, gauge invariant and the kinetic term for $\tilde{A}_{2a}$ is just canonical. Likewise, $A_{1a}$ eats the scalar $\delta X_2$ . Thus we obtain, after taking the trace over the diagonal metric factors also, the following equation of motion for all five Lorentz components of the two vector fields $A_{1a}$, $A_{2a}$,
\begin{equation} \label{vehiggs}
\partial_\rho[ \rho^3 \partial_\rho V_a] + {M^2 - (m_u-m_d)^2  \over 2} \left( {1 \over (\rho^2 +m_u^2)^2} + {1 \over (\rho^2 + m_d^2)^2 } \right)V_a = 0 \, .
\end{equation}
We note that in this example, all five Lorentz components of each gauge field lead to the same meson mass.  The $\delta X_2$ fluctuation and $A_{1a}$ field follow the same pattern and lead to further degenerate states.
From the boundary point of view, we obtain two vector mesons associated to the four Lorentz components $\mu \in \{0,1,2,3 \}$ of $A_{1a}$ and $A_{2a}$, respectively.  In addition, there are  two real scalar mesons arising from the fluctuations of the radial components $A_{1\rho}$ and $A_{2\rho}$.  
 In the mass degenerate case where the vector symmetry is preserved, $A_\rho$ is usually set to zero by a gauge transform. Here, however, this extra degree of freedom becomes physical. It is degenerate with the vector mesons, but appears as a scalar
in the gauge theory dual. In fact it is nothing other than the scalar meson that was previously described by the now eaten degree of freedom. Thus there are two vector mesons and two real scalar mesons made of $\bar{u}d$ and $\bar{d}u$.

Finally we must consider the off-diagonal complex fluctuations of $X$. These two real scalars acquire a mass squared from the potential eq.~(\ref{n2pot}) also proportional to $(m_u-m_d)^2$. Their equation of motion is again degenerate with the $\bar{u}d$ $\bar{d}u$ scalars already discussed, as well as the vector mesons in eq.~(\ref{vehiggs}). These are the same equations of motion as were numerically studied in \cite{Erdmenger:2007vj} so we do not explore the numerical solutions further here. 

In summary we have spent a considerable amount of time in this section developing the bottom up model of the ${\cal N}=2$ gauge theory in order to: 1) show how to build a bottom up model with appropriate potential and bulk gauge fields; 2) to explore and test our numerical method for fields that have mixed equations; and 3) to demonstrate the Higgs mechanism in the bulk is needed if symmetries are explicitly broken by quark masses. We will now further develop models of QCD and more exotic symmetry breaking patterns.

\section{A Bottom-Up Non-Abelian Dynamic AdS/QCD Model}
\label{sec:our model}

 Our focus in this section is to construct a non-abelian holographic model of QCD with dynamical symmetry breaking and explicit symmetry breaking masses. We will explore some of the subtleties associated with the non-abelian structures. As in the previous model we will take the basic components inherited from the D3/probe D7 system and make minimal adjustments to fit the theory to be modelled. 

The key element to describe QCD is to embed the global symmetries SU($N_f)_L \times$SU($N_f)_R$ and the symmetry breaking pattern to SU$(N_f)_V$. 

\subsection{Kinetic Terms}

    To describe the vacuum of QCD we will need to include the field $X$ that describes the chiral condensate. It naturally transforms under the chiral symmetries as $X \rightarrow U_L^\dagger X U_R$.  In additon we must include gauge fields to provide the holographic description of the sources and currents associated with the chiral symmetries. Our kinetic terms (as is familiar from the earliest AdS/QCD models \cite{Erlich:2005qh,DaRold:2005mxj}  although the factors of $r$ are adjusted since to include the backreaction of $X$ which has dimension one) are

\begin{equation}
S = \int d^5 x ~ \rho^3 \text{STr}\left( \frac{1}{r^2} (D^a X)^{\dagger} (D_a X) + \frac{1}{ g^2_5} \left(\vphantom{\frac{1}{2}} F_{L,ab}F_{L}^{ab} + (L \leftrightarrow R) \vphantom{\frac{1}{2}} \right) \right) \, .
\end{equation}

The five-dimensional coupling may be obtained by matching to the UV vector-vector correlator \cite{Erlich:2005qh}, and  is given by 
\begin{equation}
g_5^2 = \frac{24 \pi^2}{d(R)~N_{f}(R)} \, ,
\end{equation}
where $d(R)$ is the dimension of the quark's representation and $N_f$(R) is the number of flavours in that representation.
The covariant derivative is \begin{equation}
\label{eq:bu_covariant_derivative}
    D^aX=\partial^a X-\ii A^a_LX+iXA^a_R \,.
\end{equation} 

The model lives in a five-dimensional AdS$_5$ spacetime, which is given by 
\begin{align}  \label{thing2}
ds^2 = r^2 dx^2_{(1,3)} + \frac{d \rho^2}{r^2} \, ,
\end{align} 
however, again we must promote these metric elements to matrices that transform also under the chiral symmetries as $G \rightarrow U_L^\dagger G U_R$. This essentially means writing $r$ as a matrix
\begin{align}
r^2 = \rho^2 \id_{N_f} + X^\dagger X.
\end{align}
Note that formally the identity here is some combination of metric elements such as $G_{\rho \rho} G_{xx}$ that is the identity but transforms under the chiral symmetries. The STr represents that we include the metric terms in all possible positions allowed by the symmetries of the model equally. 

\subsection{Potential}

To induce dynamical chiral symmetry breaking in the model we must include a potential for the X fields which naturally takes the from 
\begin{equation} \label{BC}
V = \STr \left( A + B X^\dagger X + C (X^\dagger X)^2 + ... \right) \, ,
\end{equation}
where the coefficients may be $\rho$ dependent  (representing the entering of metric components etc of the background) and to ensure all terms are of the correct dimension. At this stage we assume they are flavour independent since any flavour breaking (including quark masses) will be generated as vevs for the bulk fields. The coefficients are therefore scalars rather than matrices. $A$ only contributes to the vacuum energy and we do not fix it.

$B$, which we call $\Delta m^2$ below, is a contribution to the mass of the X fields. To understand it's role, let's return to the abelian D7 probe computations briefly.
An example of a chiral symmetry breaking set up in the probe D7 system is  obtained by adding a world-volume baryon number magnetic field \cite{Filev:2007gb}, ${\cal B}$. This breaks supersymmetry and conformality.  The DBI action arranges to give the usual action with an effective dilaton multiplier
\begin{equation} \label{Bdil} 
e^{- \phi} = \sqrt{1+ {{\cal B}^2 \over r^4}}.
\end{equation}
The resulting equations of motion for the vacuum configuration for $L$ have solutions with $L=0$ in the UV (for large $\rho$) that bend off the $L=0$ axis in the interior.
These solutions  break the U(1)$_A$ chiral symmetry. The reason for this behaviour follows from the divergent behaviour of the dilaton factor - for example in (\cref{Bdil}) the action clearly grows as $r^2=\rho^2 + L^2 \rightarrow 0$. One can further see an instability though by expanding the dilaton factor around $L=0$, the chirally symmetric vacuum,
\begin{equation} 
e^{- \phi} = \sqrt{1+ {{\cal B}^2 \over \rho^4}} (1 - { {\cal B}^2 \over \sqrt{1+ {{\cal B}^2 \over \rho^4}} \rho^6} L^2 + \dots).
\end{equation}
The  ${\cal O}(L^2)$ term in the expansion is simply a mass term although in this case $\rho$ dependent. At small $\rho$ the mass grows until it violates the Breitenlohner Freedman bound \cite{Breitenlohner:1982jf} (this is when this contribution to the mass $\Delta m^2= -1$  since the field $L$ has intrinsic dimension one in AdS$_5$) and the $L=0$ solution becomes unbounded. In the AdS duality the mass is precisely linked to the dimension $\gamma$ of the mass and quark condensate operator that  $L$ is dual to: $\Delta m^2 = \gamma(\gamma-2)$. The instability sets in when the anomalous dimension of the quark mass $\gamma=1$ - see \cite{Alvares:2012kr} for more detailed discussion of this instability. 

The bottom-up dynamic AdS/QCD model \cite{Alho:2013dka,Erdmenger:2020flu} took inspiration from this mechanism to simply include a potential inspired by the running of $\gamma$ in the gauge theory. Here, to match the perturbative regime we set \cite{Alho:2013dka,Erdmenger:2020flu}
\begin{equation} \label{dmmu}
\Delta m^2 = - 2 \gamma, \hspace{1cm}   \gamma = {3 C_2(R) \over 2 \pi} \alpha 
\end{equation}
where we have quoted the gauge theory's one-loop running of $\gamma$ in terms of the running of $\alpha$. In previous papers we have taken the running of $\alpha$ from the two loop gauge theory result setting $\mu = r = \sqrt{\rho^2 +L^2}$. We will discuss this identification in more detail below.

The two-loop result for the running coupling in a gauge theory with multi-representational matter is given by
\begin{align}
\mu \frac{d \alpha}{d \mu} = - b_0  \alpha^2 - b_1  \alpha^3 \, ,
\end{align} 
with 
\begin{align}  
\label{running}
\begin{aligned}
b_0 &= \frac{1}{6 \pi} \left(11 C_{2}(G) - 2 \sum_{R} T(R)N_f(R) \right) \, ,\\
b_1 &= \frac{1}{24 \pi^2} \left(34 C^2_{2}(G) - \sum_{R} \left(10 C_{2}(G) + 6 C_{2}(R) \right) T(R) N_f(R) \right) \, .
\end{aligned}
\end{align}
Note, that we have written the results for Weyl fermions instead of Dirac fermions in a given representation as this is more useful in case of Composite Higgs models
\cite{Erdmenger:2020flu}.

 We now convert this logic to a bottom up model of QCD's non-abelian flavour symmetries. The base Lagrangian in the scalar sector is
\begin{equation}
S = \int d^5 x ~ \rho^3 \text{STr}\left( \frac{1}{r^2} (D^a X)^{\dagger} (D_a X) + \Delta m^2 X^\dagger X  \right).
\end{equation}
As discussed, a  key point for non-abelian extensions of the abelian case is that the metric components or equivalently $r^2$ is a matrix as in \cref{eq:matrixr}.

Let's begin by assuming $\Delta m^2$ is the flavour independent scalar quantity $B$ we introduced in the potential above. That is we make it a $\rho$ dependent function by setting $\mu = \rho$ in (\ref{dmmu}). Now in QCD the SU($N_f)_L \times SU(N_f)_R$ chiral symmetries are sufficient to diagonalize the chiral condensate matrix. We will therefore assume the vacuum state of $X$ is diagonal and real $X_0=$ diag($L_u,L_d,..)$. The $L_i$ satisfy the equations
\begin{align}
\partial_{\rho} (\rho^3 \partial_{\rho} L_i) - \rho ~ \Delta m^2(\rho) L_i = 0 \quad (i=1,\dots,N_f) \label{eq:vacuum_qcd} \, .
\end{align}
As in the abelian case though here there is a BF bound violation at small $\rho$ which can not be removed by the formation of a vev for the $L_i$. To remove this we naturally want to make the shift $\rho \rightarrow \sqrt{\rho^2 + L_i^2}$ in each equation. This we will do but it intrinsically implies that we have made $\Delta m^2$ a matrix that must be included inside the STr in the action. If one expands that matrix in powers of $X^\dagger X$ then we can see that we have effectively chosen all of the coefficients $B,C,...$ in (\ref{BC}) to return the equation of motion we desire.  The shift $\rho \rightarrow \sqrt{\rho^2 + L_i^2}$ is a well motivated choice of these parameters though. 

We solve (\ref{eq:vacuum_qcd}) with the initial conditions in the infrared (IR)
\begin{align}
L_i(\rho_{IR,i}) = \rho_{IR,i} \quad
\text{and} \quad L_i'(\rho_{IR,i}) = 0 \,.
\end{align}
In the UV one demands that
$L_i(\rho\to\infty)=m_i$. $\rho_{IR,i}$ are the scales where each quark goes on mass shell. 
In practice these IR values
are quite similar for the $u,d,s$ quarks despite possible large hierarchies of UV quark masses. In the following subsections we will consider fluctuations describing spin zero and spin one states.
We will set the corresponding boundary conditions in the IR at
$\rho_{IR} = $max$(\rho_{IR,i})$
the scale where the highest mass quark component goes on mass shell. 

We will parameterize the scalar fluctuations as  
\begin{equation}
\label{eq:bu_X}
X = e^{i\pi_a(x) T_a} \left(L+\ \sum_k \xi_k(x) T_k \right)
 e^{i\pi_a(x) T_a} \,,
\end{equation}
where in each case the generators are four orthogonal (Tr$T^aT^b={1 \over 2} \delta^{ab}$) basis matrices. 
A natural basis are the generators of SU($N_f$) plus $T_0 = \frac{1}{\sqrt{2 N_f}}  \id_{N_f}$ but we will also discuss the linear combinations ${1 \over \sqrt{2}} (T_0 \pm T_3)$ in the SU(2) example below.

Moreover, we will use the combinations
\begin{align}
 A=\frac{1}{2} \left(A_L - A_R \right)\quad \text{and}\quad V= \frac{1}{2} \left(A_L + A_R \right)
\end{align}
where $V$ corresponds to vector states and $A$ to axial vector states.

The equations of motion can be found from the abelian case by including the STr over the matrix valued components. For example, for the real scalar $\xi_k$ and the vector field which we write as $V^\mu = V^\mu_{trans} + \partial^\mu \phi_V$ we find
\begin{equation} \label{full} \begin{array}{c}
\partial_{\rho} (\rho^3 \pdr \xi_k(\rho)) - \rho \xi_j \STr \Delta m^2 (T^k)( T^j) -  \rho \xi_j \STr X_{0}  \frac{\partial \Delta m^2}{\partial L} |_{X_{0}} (T^k) (T^j) + M^2 \xi_j \STr\frac{\rho^3}{r^{4}} (T^k) (T^j)  \\ \\+\pd_\rho \STr\lrbb{\rho^3 2i\lrmb{\pi^j\lrb{(\pdr X_o^\dag)T^kT^j+(\pdr X_0^\dag T^j)(T^k)}+\pdr\pi^j(X_0T^j)(T^k)}+h.c}\\ \\
+\rho^3\pd_\rho\pi^b\STr(i(\pd_\rho X_0)T^kT^j+h.c.)
+\rho^3M^2\pi_j \STr(i{1\over r^4}(X_0T^j)(T^k)+h.c.)\\ \\ - \rho^3 M^2\phi_V^j \STr (i{1 \over r^4}   [T^j, X_0](T^k)+h.c.) -M^2\rho^3\phi_A^j\STr(i{1\over r^4}\{T^j,X_0\}(T^k)+h.c.)=0
\end{array}
\end{equation}

   \begin{equation}
   \partial_\rho (\rho^3 \partial \xi_V^a)
   - \rho^3 g_5^2 (\xi^b  \STr ({1 \over r^4}   T^a  T^b)-\phi_V^b \STr ({1 \over r^4}    [T^a, T^b]) ) = 0
    \end{equation} 
    
    For the perpendicular components of the vector gauge field we have
    \begin{align} \label{eq: eqn of motion_vector_general_param}
\partial_{\rho} (\rho^3 \partial_{\rho}V^a(\rho)) + M^2_{V} \frac{\rho^3}{r^{4}} V^a(\rho) + \rho^3 g_5^2 \STr ([X^\dagger, T^b][X,T^a]) V^b(\rho)= 0.
\end{align}
As one can see, for a generic parametrization of the field $X$ as in \cref{eq:general_param} with a non-diagonal vev there will be mixing between essentially all fields in the model. In fact we find that for the particular parametrization of $X$ in (\ref{eq:bu_X}) the fields in the vector and axial pieces of the scalar and vector remain unmixed even for a  $X$ vev that is not proportional to the identity.

\subsection{The Higgs Mechanism for the Vector Gauge Field}

When the quark masses are unequal, the vector symmetry is explicitly broken in the gauge theory. In the bulk though, the quark masses emerge in the solutions of the equations of motion for the entries in the $X$ matrix in flavour space,  and there is a vector gauge field still present. This gauge symmetry is naturally higgsed in the bulk gravity theory.

To see the Goldstone mode  consider a $N_f=2$ version of the theory with a truncated scalar potential
\begin{equation}
    {\cal L} = \rho^3 \STr \partial_\rho X^\dagger \partial_\rho X + A(\rho) \STr X^\dag X + B(\rho) \STr X^\dag X  X^\dag X 
\end{equation}
here in our model $A(\rho) = \Delta m^2(\rho)$ and $B(\rho) = {\partial \Delta m^2\over \partial \rho^2}$  with further terms in the Taylor expansion dropped. The vacuum is given by diagonal elements of $X$, $\lu, \ld$ satisfying
\begin{equation}
    \partial_\rho (\rho^3 \partial_\rho \ludn) - A \ludn - 2 B \ludn^3 = 0
\end{equation}
Now consider a fluctuation $\delta X = \left( \begin{array}{cc} 0 & \xi_2 \\ \xi_2 & 0 \end{array} \right)$ with quadratic action
\begin{equation}
    {\cal L} = \rho^2 (\partial_\rho \xi_2)^2 + A \xi_2^2 + B( 2 \lun^2 \xi_2^2 + 2 \lun \ldn \xi_2^2 + 2 \ldn^2 \xi_2^2)
\end{equation}
here we have not included space-time dependent kinetic terms because we will seek a massless solution on which they would vanish. 
The resulting equation of motion has the particular solution $\xi_2 = \lun - \ldn$. This is the Goldstone from the bulk perspective that is eaten by the vector gauge field when $\lun \neq \ldn$. In the field theory this is not a physical state because $\lun - \ldn$ does not vanish asymptotically. Nevertheless it is important to write the potential in the expanded form in (4.6) to correctly generate the equations of motion for the off-diagonal fluctuations. 

Now lets include the vector field. We can derive three equations of motion - one for the scalar $\xi_2$, 

\begin{equation} \label{one}
    \begin{aligned}
   \quad \partial_\rho\left(\rho^3 \partial_\rho \xi_2(\rho)\right)+\frac{\rho}{2}  \left[-2A- B(4 \lun^2 + 4 \lun \ldn + 4 \ldn^2) \right] \xi_2&\\
    +\frac{1}{2}
   M_{\xi_2}^2 \rho ^3 \left(\frac{1}{r_d^4}+\frac{1}{r_{u}^4}\right) \left(\xi_2(\rho)\pm(\lun-\ldn)\phi_{V_{2/1}}(\rho )\right)&=0
  \end{aligned}
\end{equation}
and two for the vector field that we write as $V^\mu= V_{\perp}^{\mu} + \partial^\mu \phi_{V}$, $\pd_\mu V^\mu_\perp=0$, $V_\rho=0$ - the first is the equation of motion for $V^\mu$ with this form substituted and the other the direct equation for $\phi_V$ 
\begin{equation} \label{two}
    \begin{aligned}
      \partial_\rho\left(\rho^3 \partial_\rho\phi_{V_{1/2}}(\rho)\right)\pm\frac{\rho^3 g_5^2}{8}\left({1\over r_u^4}+{1\over r_d^4}\right)\left( \lun-\ldn\right) \lrb{\xi_2(\rho)\mp(\lun-\ldn)\phi_{V_{1/2}}(\rho )} &=0\\
   4M^2\pdr\phi_{V_{1/2}}(\rho)\mp g_5^2(\lun- \ldn)\pdr\xi_2(\rho)\pm g_5^2 \xi_2(\rho) \pdr(\lun-\ldn)&=0
  \end{aligned}
\end{equation}
There are really only two equations with one redundant - for example substituting the bottom equation in eq.~(\ref{two}) into the top one lead to eq.~(\ref{one}). 

Were one to include higher order terms in the expansion (STr$(X^\dagger X)^n$) this same Higgs mechanism and consistency holds - in a sense eq.~(\ref{two}) implicitly contains the information of the potential in eq.~(\ref{one}) through the solutions $L_i$.

This Higgs mechanism in the bulk is rather elegant since it shows how explicit breaking in the gauge theory translates to the bulk gauge symmetry. However, unfortunately when in eq.~(\ref{eq:vacuum_qcd}) we impose the running of $\Delta m^2$ at the level of the equation of motion rather than in the Lagrangian  we spoil these consistency conditions. Then if we use (\ref{two}) we do not get numerical results that give degeneracy of the $\xi^3$ and $\xi^{1/2}$ as the mass splitting vanishes. Instead of using the full (\ref{one}), in the numerics below, for small mass splittings, we will ignore the vector field mixing (the last term in (\ref{one}) of vanishes as $\lun=\ldn$) and use the $\xi_2$ equation of motion
\begin{equation} \begin{array}{c}
   \partial_\rho\left(\rho^3 \partial_\rho \xi_2(\rho)\right)+\frac{\rho}{2}  \left[-\Delta m_{u}^2-\Delta m_{d}^2-L_u(\rho ) \Delta
   m_{u}^{2\prime}-L_d(\rho ) \Delta m_{d}^{2\prime}\right]\xi_{1/2}(\rho )\\ \\
    +\frac{1}{2}
   M_{\xi_{1/2}}^2 \rho ^3 \left(\frac{1}{r_d^4}+\frac{1}{r_{u}^4}\right) \xi_2=0
   \end{array}
  \end{equation}
which is consistent with the substitution in  (\ref{eq:vacuum_qcd}) we have made. 

     
    
Using these simplifications, we will now consider some particular phenomenologically interesting cases.

\subsection{Scenario 1 - $N_f$ Equal Masses}

\begin{figure}
    \subfigure[Embedding vs. quark masses.]{\includegraphics[width=0.48\textwidth]{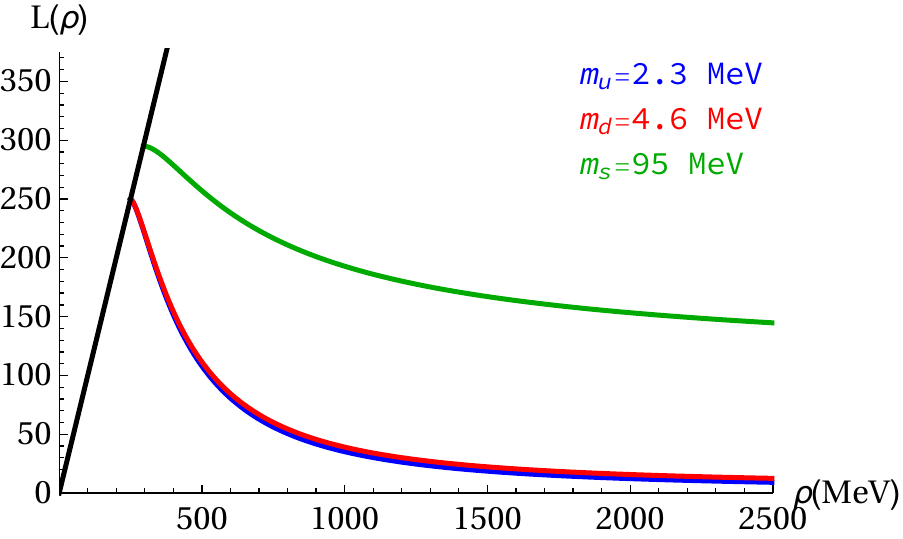}}
    \label{fig:L_vs_M}
    \hfill
    \subfigure[Embedding with $\kappa$.]{\includegraphics[width=0.48\textwidth]{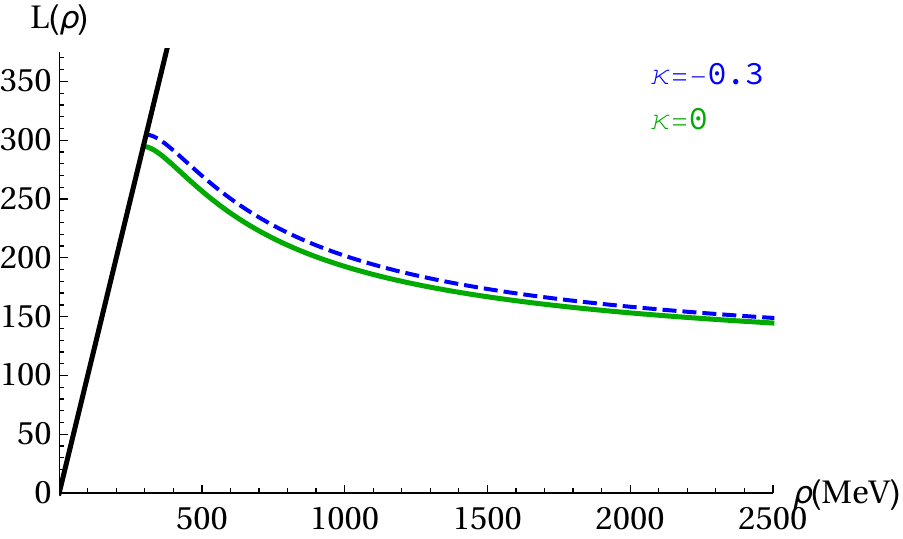}}
    \label{fig:L_vs_kappa}
      \caption{Left plot: different embeddings for various  quark masses.  The quark masses are $m_u=2.3$ MeV, $m_d=4.6$ MeV and $m_s=95.5$ MeV respectively, after matching the $\rho$ meson mass to 775 MeV.  Right plot:  the dashed line shows the effect of a double trace term with $\kappa=-0.3$ (dashed line) and $m_s=95.5$ MeV compared to $\kappa=0$ (full line).}
      \label{fig:L}
\end{figure}

For the case of a diagonal quark mass matrix the vacuum structure of the theory breaks into $N_f$ copies of the $N_f=1$ case. The equation of motion for each real diagonal component of $X$ is
\begin{align}
\partial_{\rho} (\rho^3 \partial_{\rho} L) - \rho ~ \Delta m^2 L = 0 \label{eq: vacuum qcd} \, ,
\end{align}
where here $\Delta m^2$ is simply a scalar function corresponding to any one of its equal diagonal components. We have dropped the indices here to simplify the notation.
We insert the $r$ dependent $\Delta m^2$ here at the level of the
equation of motion so that we are precisely using it to set the
running anomalous dimension. We plot the solution for a variety of
common quark masses in the left plot of \cref{fig:L}. 
In the UV, the curve will flatten to the given quark masses, as can be seen from the $u$ and $d$ quark embeddings. The strange quark embedding will eventually tend to 95 MeV, which is not shown explicitly in this plot.
We set the scale with the rho meson mass.
The solution of eq.~\eqref{eq: vacuum qcd} is denoted by $L_0$ below.

We now discuss the fluctuations (mesons) of the theory. Let us begin the scalar sector where it is sensible here to discuss the isospin triplet, $\xi^k$, and isospin scalar, $\sigma$, states. The kinetic terms for the $\sigma, \xi_k$ fields are simply quadratic (any linear terms cancel when evaluated on the solutions of the equation of motion) and separately follow the basic trace algebra 
\begin{equation} 
\Tr \left(\sigma {1\over \sqrt{2N_f}} \id_{N_f} + \sum_k \xi_k(x) T_k \right)^2 = {1\over 2}\left(\sigma^2 + \sum_k \xi_k^2 \right).
\end{equation}
Since the vev of $X$ is proportional to the identity we can
move the $\Delta m^2$ and metric factors outside of the STr in eq.~(\ref{full}). The trace treats $\sigma$ and $\xi_k$ on an equal
footing - they will therefore be degenerate. 

The equation of motion, consistent with the truncation in \cref{eq: vacuum qcd},  for the $\sigma$ fluctuation reads
\begin{equation} \label{eq: eqn of motion_scalar}
\partial_{\rho} (\rho^3 \partial_{\rho} \sigma(\rho)) - \rho \Delta m^2 \sigma(\rho) - \rho L_{0}(\rho) \sigma(\rho) \frac{\partial \Delta m^2}{\partial L} |_{L_{0}} + M^2 \frac{\rho^3}{r^{4}} \sigma(\rho) = 0.
\end{equation}

The $N_f^2$ vector-mesons are obtained from fluctuations of the gauge fields $V={1\over 2}(A_L+A_R)$ around the vacuum. They couple to $X$ via a commutator which is zero for $X \propto \id_{N_f}$ so their equations are also governed simply by the quadratic $F^2$ kinetic term. 
They are all degenerate and satisfy the equation of motion
\begin{align} \label{eq: eqn of motion_vector}
\partial_{\rho} (\rho^3 \partial_{\rho}V(\rho)) + M^2_{V} \frac{\rho^3}{r^{4}} V(\rho) = 0.
\end{align}

 The axial-vector meson gauge field in the bulk enters the covariant derivative for the field $X$, coupling as an anti-commutator. The result is that a $\rho$ dependent mass term proportional to $L_0^2$ forms. In  choosing the $A_\rho =0$ gauge and decompose the axial-vector as $A_\mu = A_{\mu\perp} + \partial_\mu \phi$, with $\pd_\mu A^\mu_\perp=0$, one observes a Higgs mechanism. The action is of the form
\begin{equation}
S= \int d^4x d \rho \left[ -{\rho^3 \over 4 g_5^2} F^a_A F^a_A +  {L_0^2 \rho^3} (\partial \pi - A^a)^2 \right],
\label{eq:lag_axial}
\end{equation}
where we have suppressed the space-time indices.
One arrives at the equation of motion for the $N_f^2$ axial mesons which are degenerate
\begin{equation}  \label{eq: eqn of motion_axial}
\partial_{\rho} (\rho^3 \partial_{\rho} A(\rho)) - g_5^2 \frac{\rho^3 L^2_{0}}{r^2} A(\rho) + \frac{\rho^3 M^2_{A} }{r^{4}} A(\rho) = 0\, . 
\end{equation}

The $\phi^a$ and $\pi^a$ fields (the phases of $X$)  mix to describe the pion - we have the two equations of motion
\begin{equation} 
\partial_\rho (\rho^3 L_0^2 \partial_\rho \pi^a) - {L_0 \rho^3 q^2 \over (\rho^2+L_0)^2} (\pi^a - \phi^a)=0
\end{equation} 
\begin{equation} 
q^2 \partial_\rho (\rho^3  \partial_\rho \phi^a) - {L_0 \rho^3 q^2 \over (\rho^2+L_0)^2} (\pi^a - \phi^a)=0.
\end{equation} 
The difference of these two gives a total derivative that can be integrated and the constant determined to be zero at large $\rho$ so
\begin{equation} q^2 \partial_{\rho} \phi- g_5^2 L_0^2 \partial_{\rho} \pi = 0.
\end{equation}
The solutions of these equations have been previously studied in \cite{Erdmenger:2020flu} and generate $N_f^2$ massless pions in the zero quark mass limit and display a Gell-Mann-Oakes-Renner relation at finite quark mass. We present numerical computations of the meson masses in the next section where we also include $1/N$ effects.

\subsection{Scenario 2 - Two Equal-Mass Quarks and $1/N$ Effects}

In Scenario 1 above,  all the terms we have lead to degeneracy between the $N_f^2$ states in the vector and scalar meson sectors. Generically these states split into a $N_f^2-1$ dimensional representation of SU($N_f$) and a singlet.  To include such splitting we must add for example additional terms  to our scalar potential which must be invariant under the symmetries. To see the key point it is useful to explicitly compute the operator 
\begin{equation} 
\left(\Tr X^\dagger X\right)^2 = {1\over 2}\left[(L_0 + \sigma)^2 + \sum_k \xi_k^2\right]
\simeq 2 L_0^2 \xi_k^2 + 6 L_0^2 \sigma^2 + \dots .
\end{equation}
Clearly these terms break the  degeneracy between $\xi_k$ and $\sigma$.
We add the double trace term $\kappa \Tr[X^\dagger X]$ to the Lagrangian to exemplify the effect. It will also change the
equation of emotion for the embedding which now reads as
\begin{equation}
  \label{eq:double_trace_vac}
  \partial_\rho\left(\rho^3\partial_\rho L_0(\rho)\right)-\rho\Delta m^2(\rho)L_0(\rho)-\kappa L_0(\rho)^3\rho=0 \,.
\end{equation}
The numerical effect of such a contribution is however small as can be seen from the righthand plot of \cref{fig:L} where we show the case relevant for the strange quark. In case of smaller masses, e.g.\ for the $u$- and $d$-quarks, the cases $\kappa=0$ and $\kappa=-0.3$ can hardly be distinguished.
\begin{figure}[t]
  \centering
  \includegraphics[width=10cm, height=6cm]{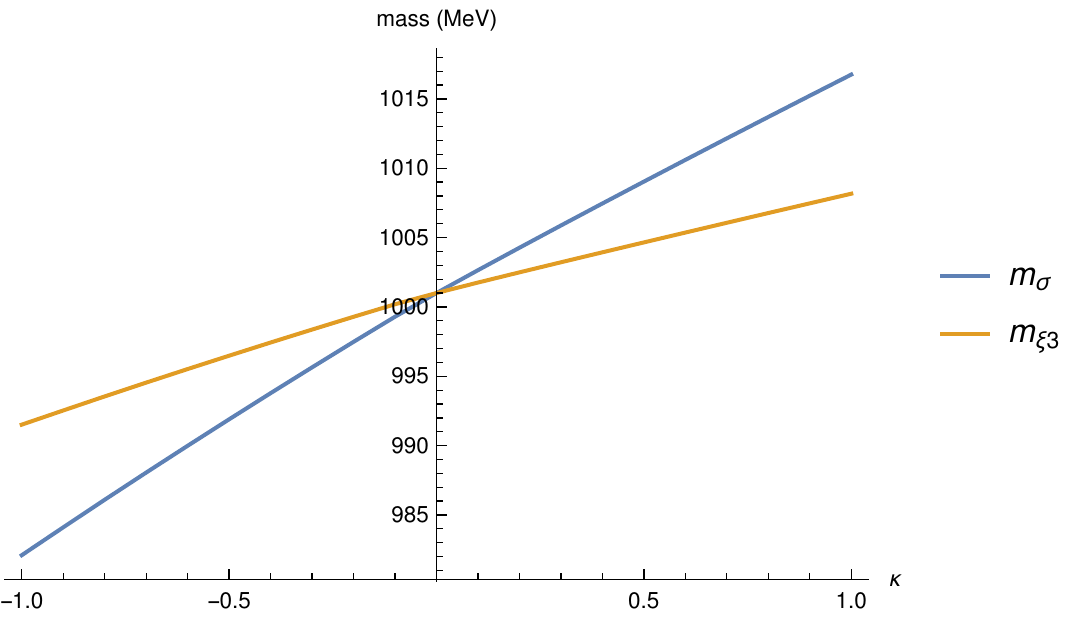}
  \caption{Dependence of the scalar masses on $\kappa$ when quark mass $m_q=2$ MeV. $\kappa$ is the coupling of the double trace term introduced in \cref{eq:scalar_add_double_trace}, which gives a small splitting in the singlet and triplet scalar masses $m_\sigma$ and $m_{\xi_3}$.}
  \label{fig:kappa}
\end{figure}
This operator affects in particular the equations of motion of $\sigma$ and $\xi_3$, they are changed to
\begin{equation}
\label{eq:scalar_add_double_trace}
  \begin{aligned}
    &\partial_{\rho} (\rho^3 \partial_{\rho} \sigma(\rho)) - \rho (\Delta m^2) \sigma(\rho) - \rho L_{0}(\rho) \sigma(\rho) \frac{\partial \Delta m^2}{\partial L} |_{L_{0}} + M_\sigma^2 \frac{\rho^3}{r^{4}} \sigma(\rho)-3\kappa L_0^2\sigma(\rho)\rho = 0\\
    &\partial_{\rho} (\rho^3 \partial_{\rho} \xi_3(\rho)) - \rho (\Delta m^2) \xi_3(\rho) - \rho L_{0}(\rho) \xi_3(\rho) \frac{\partial \Delta m^2}{\partial L} |_{L_{0}} + M_{\xi_3}^2 \frac{\rho^3}{r^{4}} \xi_3(\rho)-\kappa L_0^2\xi_3(\rho)\rho = 0.
  \end{aligned}
\end{equation}
Solving the equations numerically, we find a dependence of the
scalar masses on the factor $\kappa$, as shown in
\cref{fig:kappa}. 
Double trace terms in QCD are expected to be suppressed by $1/N^2$ so we have chosen a fairly narrow range of $\kappa$ values  in \cref{fig:kappa}.

We solve the meson masses from \cref{eq:scalar_add_double_trace} using the shooting method, the spectrum is listed in \cref{tab:same_mass_spectrum}. 
 The vector $\rho$ meson mass is used as an input parameter to read out physical masses of the other lowest meson states. We find that most of the masses lie within 3\% range with respect to the data. The pion ($\pi^0$) mass is very sensitive to the quark mass $m_q$, this explains the larger deviation comparing with the others. Notice that, adding double trace term in the same mass scenario does produce a realistic mass splitting between the scalar singlet ($f_0(980)$) and triplet ($a_0(980)$) mass.

 \begin{table}[h]
    \centering
    \begin{tabular}{c|c|c|c}
    \hline
     Observables&QCD [MeV]& $U(N_f=2)$ [MeV] & Deviation\\
     \hline

       $M_\rho(770)$ &$775.26\pm0.23$ &775* & fitted\\
        $M_{a_1(1260)}$ & $1230\pm40$ &1194 & 3\%\\
      $ M_{f_0(980)} $& $990\pm20$ &994* & $<1$\%\\
      $M_{a_0(980)}$ & $980\pm20$ &997 & 2\%\\
        $\pi^0$ & $134.9768\pm0.0005$ & 117 & 14\%\\ 
        \hline
    \end{tabular}
    \caption{Meson masses of the lowest lying states in the U(N$_f$) model with equal quark masses. The
      $\rho$ meson mass is fixed to 775 MeV and we have set $m_q=m_u=2.3$ MeV and $\kappa=-0.3$ (therefore the asterisk next to 994.). The QCD masses are taken from the PDG \cite{Workman:2022ynf}. The mass difference in $f_0(980)$ and $a_0(980)$ is introduced by the double trace term with the coupling $\kappa$.  The $\pi^0$ mass is very sensitive to the quark mass, this explains the large deviation.}
    \label{tab:same_mass_spectrum}
\end{table}

\subsection{Scenario 3 - $N_f=2$ Split Masses}  
\label{sec:Nf2_split}
In this section, we discuss the case of unequal masses for the different quark flavours. To start with, we consider a two flavour theory with different quark masses
($m_u\ne m_d$) to exemplify the features of
the non-abelian DBI action. 
Here we neglect the  additional contributions in QCD from the electromagnetic interactions.
The $U(N_f)_L \times U(N_f)_R$ symmetry can be used to find a basis in which the mass matrix is real and diagonal. To begin with, consider only including single trace terms in the potential for the mass splitting case. At this stage, the vacuum is expected to also be diagonal - the matrix structure simply falls apart into two copies of the one flavour case but here we set different IR boundary conditions on the two $L_0$s to represent the different UV masses - we can call the two solutions $L_u$ and $L_d$
\begin{equation}
\label{eq:vac_eoms_separated}
    \partial_\rho \left(\rho^3 \partial_\rho L_i\right)-\Delta m_{i}^2\rho L_i=0,\quad i=u,d.
\end{equation}
The vacuum now preserves a U(1) vector symmetry in each of the $u$ and $d$ quark sectors. In this large $N_c$ limit with no multi-trace terms the mesons made of $\bar{uu}$ or $\bar{d}d$ are unmixed mass eigenstates (the $\sigma $ and $\xi_3$ states are not mass eigenstates)- thus one just repeats the two separate sectors with different $L_0$. 

The mixed $\bar{u} d$ states see the mass splitting though.
After taking the symmetrised trace, we find the equations of motion are sorted into two classes. We take here the vectors as an example, the equations of motion for other fluctuations are listed in \cref{appendix:Nf=2_no_str}. The off-diagonal  $V_{1,2}$ are  dual with the meson states consists of two flavours of quarks
\begin{equation}
    \label{eq:Nf=2_TR_eoms_v12}
    \begin{aligned}
   V_{1/2}:\quad &\partial_\rho\left(\rho^3 \partial_\rho V_{1/2}(\rho)\right)+\frac{\rho^3}{8}\left({1\over r_u^4}+{1\over r_d^4}\right)\left[4 M_{V_{1/2}}^2-g_5^2\left( L_u(\rho )-L_d(\rho )\right)^2 \right]V_{1/2}(\rho ) =0.
  \end{aligned}
\end{equation}
In writing these equations, we have defined $V_i^\mu=V^\mu_{\perp,i}+\partial_\mu\phi_{V,i}$, $\pd_\mu V^\mu_\perp=0$ and taking the $V_\rho=0$ gauge, and here we have concentrated on the transverse pieces. We observe a Higgs mechanism since the longitudinal piece $\phi_{v_i}$ mixes with the scalars, see \cref{eq:Nf=2_TR_eoms_xi1}. 
The masses for the off-diagonal vector and axial-vector excitations can be solved using the usual procedure with the boundary conditions
\begin{equation}
\label{eq:off_diag_bc}
    V_i(L_d)=1,\quad  \partial_\rho V_i(L_d)=0,\quad V_i(\rho_{UV})=0,\quad i=1,2,
\end{equation}
i.e. we shoot from the IR and solve for the mass $M_{f_i}$ such that the field $f_i$ vanishes in the UV. 

\begin{figure}
    \subfigure[ $\pi^\pm$ vs. $\D m_q$ .\label{fig:dm_vs_mpi}]{\includegraphics[width=0.48\textwidth]{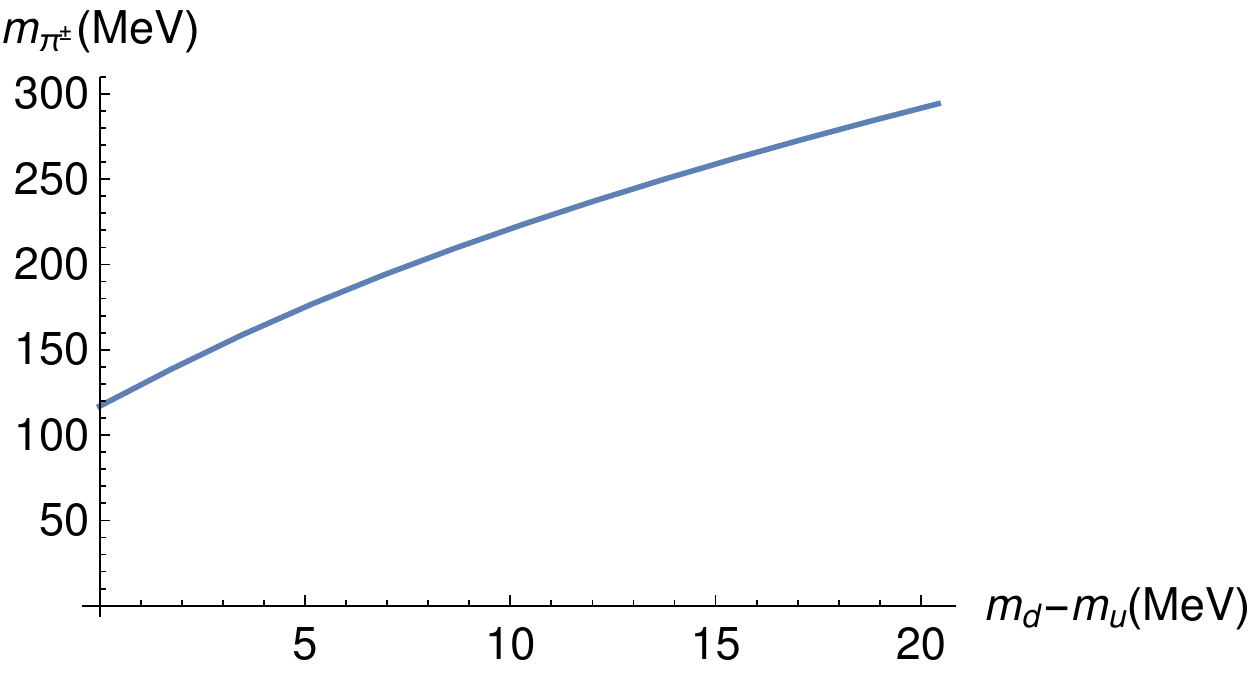}}   
    \hfill
    \subfigure[The diagonal scalar fields vary with $\kappa$.\label{fig:split_mass_with_k_scalar_masses}]{\includegraphics[width=0.48\textwidth]{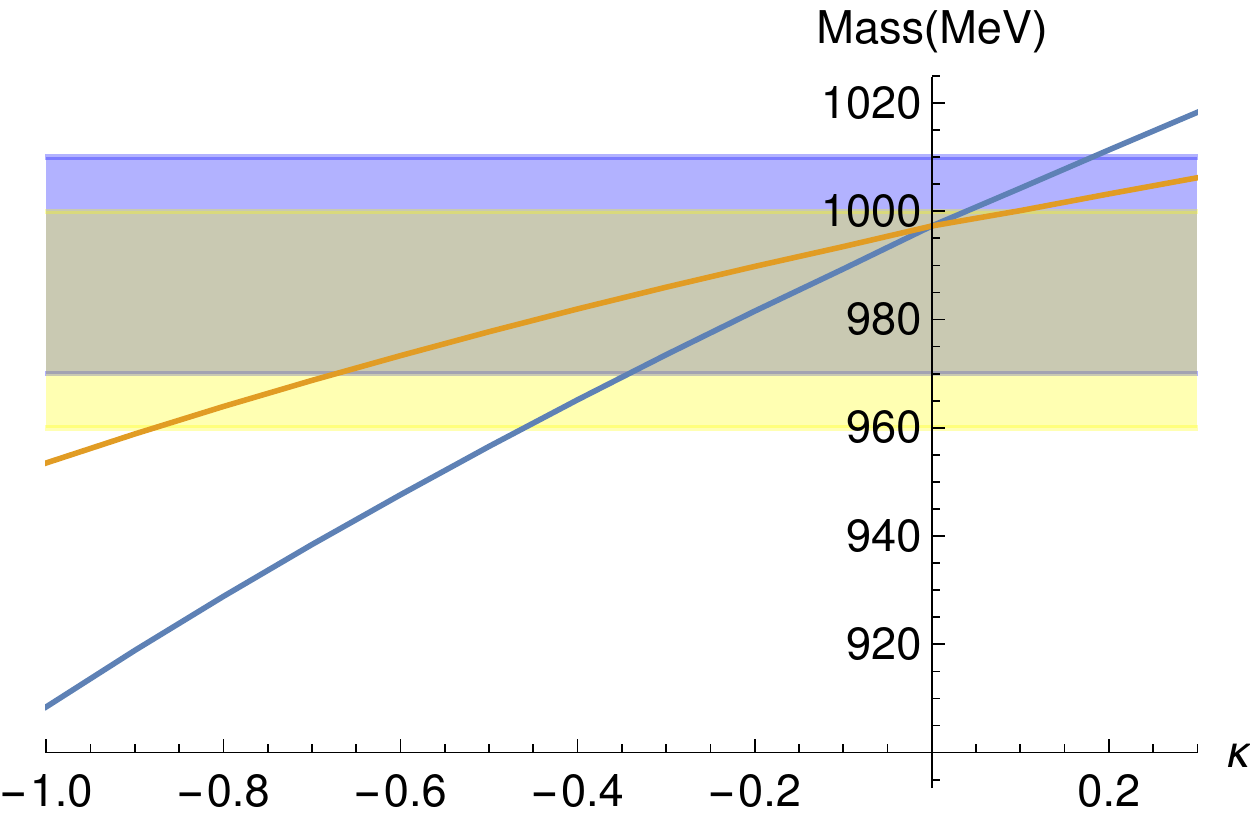}}
      \caption{Left plot: The $\pi^\pm$ mass increases with the growing quark mass splitting at fixed $m_u$ and $\kappa=0$. Right plot:The scalar masses in the presence of $\kappa\rho\Tr(X^\dagger X)^2$ in the split mass case. The shaded blue area marks the valid data range of the $f_0(980)$, and the yellow area marks that of the $a_0(980)$. This indicates a range $\kappa\in [-0.4,0]$.}
      \label{fig:split}
\end{figure}

Due to the Higgs mechanism, the off-diagonal scalars have coupled equations of motion. For example, the scalar $\xi_2$ is coupled to the longitudinal $\phi_{V_1}$. One can set the boundary condition 
\begin{equation}
    \label{eq:pi-phi_bc}
    \phi_{v,i}(IR)=1,\quad \phi_{v,i}'(IR)=0,\quad \xi_2(IR)=b, 
\end{equation}
where $b$ is a free shooting parameter as discussed in section \ref{sec:diagonal_states}. The IR scale is set by whichever of $L_u$ or $L_d$ terminates at the highest $\rho$. Again, finding the solutions that vanish for both fields in the UV gives the corresponding mass and the boundary condition $b$. However, we find that taking the limit that $\phi_{V}$ and $\phi_A$ are small compared to other fluctuations so they can be neglected is a good approximation. The resulting masses are shown in \cref{tab:Nf=2_masses}. They are very close to the triplet states shown in \cref{tab:same_mass_spectrum} due to the small physical mas splitting. After introducing the mass splitting, we see the deviation of $\pi^{\pm}$ from $\pi^0$. In \cref{fig:dm_vs_mpi} we show how the $\pi^\pm$ mass depends on the quark mass splitting.

\begin{table}[h]
    \centering
    \begin{tabular}{c|c|c|c}
    \hline
     Observables&QCD [MeV]& $U(N_f=2)$ [MeV] & Deviation\\
     \hline

       $M_\rho(770)$ &$775.26\pm0.23$ &775* & fitted\\
        $M_{a_1(1260)}$ & $1230\pm40$ &1196 & 3\%\\
      $M_{a_0(980)}$ & $980\pm20$ &998 & 2\%\\
        $\pi^\pm$ & $139.57039\pm0.00017$ & 146 & 2\%\\ 
        \hline
    \end{tabular}
    \caption{Meson masses of the lowest lying states in the U(N$_f$) model with unequal quark masses, $m_u=2.3$ MeV and $m_d=4.6$ MeV. The $\rho$ meson mass is fixed to 775 MeV. The QCD masses are taken from the PDG \cite{Workman:2022ynf}. Notice the deviation of pion from the initial $\pi^0$ state to the $\pi^\pm$ state.}
    \label{tab:Nf=2_masses}
\end{table}

Finally we can add in a double trace term $\kappa\rho\Tr(X^\dag X)^2$ in addition to mass splitting. The former favours a basis where $\sigma$ and $\xi_3$ are the mass eigenstates whilst the latter prefers the $\bar{u} u$ and $\bar{d}d$ basis. In this case the true mass eigenstates are a mixture in either basis and one must solve fully coupled equations using the methodology of section \ref{sec:diagonal_states}.

Firstly the presence of the double trace term changes the vacuum dynamics governed now by
\begin{equation}
    \label{eq:vac_eom_with_kappa}
    \begin{aligned}
       &\partial_\rho\left(\rho^3\partial_\rho L_u (\rho) \right)-\Delta m_u(\rho)\rho L_u(\rho)-2\kappa L_u(\rho)^3\rho-2\kappa L_u(\rho)L_d(\rho)^2=0\\
       &\partial_\rho\left(\rho^3\partial_\rho L_d(\rho)\right)-\Delta m_d(\rho)\rho L_d(\rho)-2\kappa L_d(\rho)^3\rho-2\kappa L_d(\rho)L_u(\rho)^2=0\\
    \end{aligned}
\end{equation}
and for example give corrections to the diagonal scalar excitations: 
\begin{alignat}{1}
    \nonumber
    &\partial_\rho(\rho^3\partial_\rho\sigma_u)+\rho\left(-\Delta m_u(\rho)-L_u(\rho)\Delta m_u'(\rho)-2\kappa\left(3L_u(\rho)^2+L_d(\rho)^2\right)+{\rho^2 M^2\over r_u^4}\right)\sigma_u(\rho)\\
    & -4\kappa\rho L_u(\rho)L_d(\rho)\sigma_d(\rho)=0\\
    \nonumber
    & \partial_\rho(\rho^3\partial_\rho\sigma_d)+\rho\left(-\Delta m_d(\rho)-L_d(\rho)\Delta m_d'(\rho)-2\kappa\left(+3L_d(\rho)^2+L_u(\rho)^2\right)+{\rho^2M^2\over r_d^4}\right)\sigma_u(\rho)\\
    &-4\kappa\rho L_u(\rho)L_d(\rho)\sigma_u(\rho)=0
\end{alignat}
In \cref{fig:split_mass_with_k_scalar_masses} we show the two mass eigenstates as a function of $\kappa$.

Note we have not discussed the mass of $\eta'$, which would be close to 250 MeV in this approximation, since we have not included the effect of the chiral anomaly. 

\subsection{Scenario 4 - $N_f=3$ Split Masses, $m_u=m_d \ll m_s$}
\label{Nf=3}
In this section we  extended the  description to
three quark flavours. The main effect on the spectrum we investigate is from the mass difference between the $s$-quark and the quarks of the first generation. Thus we take here the approximation $m_u=m_d\ll m_s$.
The vacuum equations of motion  take the same form as in \cref{eq:vac_eoms_separated}, and the  three embeddings follows $L_u=L_d\ll L_s$ in the UV. 

In this approximation the isospin $SU(2)$ subgroup of the $SU(3)$ flavour group is conserved. In the neutral sector one has  a mixing between the $SU(3)$ singlet fields and the one corresponding to the 8th component of $SU(3)$. In case of vector mesons these correspond to the $\omega$ and $\phi$. In this case we do have quite some experimental information to rely on. For example  chapter 5 of \cite{Amsler:2018zkm} teaches that there is an ideally mixed case where one mass eigenstates corresponds to an $s\bar{s}$ and the second one to a $(u\bar{u}+d\bar{d})/\sqrt{2}$ combination. This implies that actually in this case the isospin $SU(2)$ gets enlarged to a $U(2)$. As a test we have used this information and define the states 
\begin{equation}
    \label{eq:N_f=3_redef_fields}
    f_{u+d}:={\sqrt{2\over 3}}f_0+{1\over \sqrt{3}}f_8,\quad f_s:={1\over \sqrt{3}}f_0-{\sqrt{2\over 3}}f_8.
\end{equation}
We denote here any fluctuation by $f$ and the indices $u+d$ and $s$ are chosen according to composition $(u\bar{u}+d\bar{d})/\sqrt{2}$ and  $s\bar{s}$, respectively.
In this approximation we set the $m_\omega = m_\rho$ for the input and obtain $m_\omega$ as a prediction which agrees quite well with the experimental value, see \cref{tab:Nf=3_masses}. In this way the equations of motion decompose into three groups:
$(u+d,1,2,3)$, $(4,5,6,7)$ and $s$, corresponding to
the adjoint, the fundamental and a singlet of $U(2)$, respectively. The corresponding equations of motion have the same structure as in the $N_f=2$ case and we summarize in  \cref{app:Nf=3} how those two cases are related. Using the same procedure as before, we obtain the various meson masses listed in \cref{tab:Nf=3_masses}.
We see that the results for the vector mesons agree quite well with the experimental data. This is a consequence of the fact, that in both sectors the ideal mixing between the singlet and the octet state is realized. In the case of the axial vectors, a sizable deviation exists which can be understood using the technique of the QCD conformal partial wave expansion \cite{Yang:2007zt}.
This explains the larger deviations observed in this sector.  In the scalar section, the $f_0(1370)$ shows an even larger deviation, however, these states that can be $\bar{q}q$, glueball or even pion molecules are hard to identify. In the case of the pseudoscalars, theoretically we can only compute the values for the pions and kaons which again agree quite well with the data. The $\eta'$ mass are shown to complete the full analysis. This value should have a somewhat large deviation given that we didn't include the chiral anomaly.

\begin{table}[t]
  \centering
  \begin{tabular}[h]{c|c|c|c}
    \hline
    Observables &  QCD [MeV] & $N_f=3$ - split masses [MeV] & Deviation\\
    \hline
     $M_{\rho(770)}$, ${\omega(782)}$ & $775.26\pm0.23$ & 775*& fitted\\
    $M_{K^*(892)}$ & $891.67\pm0.26$ & 1009& 12\%\\
    $M_{\phi(1020)}$ &$1019.461\pm0.016$ & 1048& 3\%\\ \hline
    $M_{a_1(1260)}$, $M_{f_1(1285)}$ & $1230\pm40$ & 1104 & 11\%\\
    $M_{K_1(1400)}$ &$1403\pm7$ & 1377 & 2\%\\
    $M_{f_1(1420)}$ &$1426.3\pm0.9$ & 1713 & 18\%\\\hline
    $M_{a_0(980)}$, $M_{f_0(980)}$ & $980\pm20$ & 929 &5\% \\
    $M_{K_0^*(700)}$ & $845\pm17$ &876 & 4\%\\
    $M_{f_0(1370)}$&  1370 & 970& 34\%\\ \hline
    $M_{\pi}$ & $139.57039\pm0.00017$ & 139 & 1\%\\ 
    $M_{K}$& $497.611\pm0.013$ &584 & 16\%\\
    $M_{\eta'}$ & $957.78\pm0.06 $ & 791 & 19\%\\
    \hline
  \end{tabular}
  \caption{Meson masses in the three flavour case compared with the experimental data \cite{Workman:2022ynf}. We have fixed the masses for the vector bosons $\rho$ and $\omega$ and calculated the masses for the axial vectors, the scalars and the pNGBs. 
  The quark masses used are $m_u=m_d=3.1$ MeV and $m_s=95.7$ MeV. }
  \label{tab:Nf=3_masses}
\end{table}

\section{Other Symmetry Breaking Patterns \label{Other_section}}

We have concentrated on aspects of QCD physics above to exemplify the structure of our non-abelian holographic model. It is straight forward though to generalize it to other symmetry breaking patterns - we discuss briefly two such patterns here.

\subsection{SU($2N_c) \rightarrow$ Sp($2N_c$)}

This symmetry breaking pattern can emerge in theories where there is a gauge invariant, Lorentz singlet, bi-quark operator that can condense. For example, in SU(2) (or generically Sp($2N_c$) gauge theory where the fundamental and anti-fundamental representation are identical) if we include $N_f$ Dirac quarks then there is an SU($2N_f$) flavour symmetry on the Weyl spinors. Biquark states are anti-symmetric in colour, and a Lorentz singlet is antisymmetric in spin so the state can only form in the anti-symmetric representation of flavour by the Pauli Exclusion Principle. 

In the holographic model we therefore write the field $X$ as an $2N_f \times 2 N_f$ matrix  but restricted to anti-symmetric generators. It transforms under the flavour group as $F^T X F$. The natural vacuum expectation value in the vacuum can be placed in the form
\begin{equation} 
X = \ii \left( \begin{array}{cc} 0 & \id_{N_f} \\ - \id_{N_f} & 0 \end{array} \right)
\end{equation}
which manifestly breaks the global symmetry group to Sp($2 N_c$). If mass terms are included then one must consider the vacuum in 2$\times$2 blocks with a function $L_i(\rho)$ in each case satisfying the abelian equation we have seen above and with solution asymptoting to the $i$th quark mass.

The model must also include in the bulk a gauge field for the SU($2N_f$) global symmetry of the gauge theory.  The fluctuation equations follow naturally as in the QCD case but with restrictions to the anti-symmetric generators. We leave an explicit example to future work.

\subsection{SU($2N_c) \rightarrow$ SO($2N_c$)}

This symmetry breaking pattern can emerge also in theories where there is a gauge invariant, Lorentz singlet, bi-quark operator that can condense. For example, in SO($N_c$) gauge theory where the fundamental and anti-fundamental representation are identical, if we include $N_f$ Dirac quark then there is an SU($2N_f$) flavour symmetry on the Weyl spinors. Biquark states are symmetric in colour, and a Lorentz singlet is antisymmetric in spin so the state can only form in the symmetric representation of flavour by the Pauli Exclusion Principle. 

In the holographic model we therefore write the field $X$ as an $2N_f \times 2 N_f$ matrix  but restricted to symmetric generators. It transforms under the flavour group as $F^T X F$. The natural vacuum expectation value in the vacuum can be placed in the form
\begin{equation} 
X = \left( \begin{array}{cc} 0 & \id_{N_f} \\ \id_{N_f} & 0 \end{array} \right)
\end{equation}
which manifestly breaks the global symmetry group to SO($2 N_c$). If mass terms are included then one must consider the vacuum in 2$\times$2 blocks with a function $L_i(\rho)$ in each case satisfying the abelian equation we have seen above and with solution asymptoting to the $i$th quark mass.

The model must also include in the bulk a gauge field for the SU($2N_f$) global symmetry of the gauge theory.  The fluctuation equations  follow naturally as in the QCD case but with restrictions to the symmetric generators. Again we leave explicit examples to future work.

\section{Conclusion and Outlook}
\label{conclusion}

 In this paper we have worked through a number of implications of holographic models' descriptions of non-abelian flavour symmetry. 

The non-abelian structure only seriously manifests when one includes explicit breaking of the non-abelian symmetry. The non-abelian flavour symmetry is necessarily a gauge symmetry in the bulk. In particular we have been keen to make manifest the bulk Higgs mechanism that results from the inclusion of mass terms that break a global symmetry in the dual field theory. Although these are explicit symmetry breaking in the field theory in the bulk gravitational description the masses arise as solutions of the field equations and the breaking is spontaneous. In section \ref{sec:n2 model} we displayed this Higgs mechanism in a bottom up dual of a supersymmetric theory without quark condensates. In section \ref{sec:our model} we showed it at play in a bottom up model of QCD including quark condensates. We have combined for this purpose the  AdS\//Yang Mills model of \cite{Erdmenger:2020flu} with the top-down inspired non-ablian DBI model of ref.~\cite{Erdmenger:2007vj}. This is a model with N$_f$ branes where the corresponding brane embedding depends on the respective quark mass. 

Another key computational element that arises in the study of these cases is the difficulty of, a priori, guessing the mass eigenstate basis for the meson fields. In section      \ref{sec:diagonal_states} we developed a numerical method to find these mass eigenstates even when generically there are many mixed fields. Although this method is already in the literature we presented here an analytically solvable model (of an ${\cal N}=2$ supersymmetric theory) that allowed us to verify its veracity when making explicit changes of variables. In section \ref{sec:our model}, in the context of QCD, we used this method to compute the mass eigenstates in models with both $u$ $d$ quark mass splitting and a $1/N$ suppressed, multi-trace term that splits the isospin singlet $\sigma$ and isospin triplet $\xi^a$ mesons. Away from the large $N$  limit the mass eigenstates here are neither the $\sigma$ and $\xi^3$ nor the $\bar{u}u$ and $\bar{d}d$ states. 

We have used these methods here to fully include the effects of the non-abelian flavour structure in the calculation of the masses of bosonic QCD bound states. We calculated the mass spectrum of QCD with N$_f$ light quarks including different quark masses and found good agreement with the observed spectrum in the three flavour case.

We have briefly discussed how the framework could be extended to models with SU(2$N_f)\rightarrow Sp(N_f)/SO(N_f)$ flavour symmetry breaking patterns. Such breaking patterns are important for composite Higgs models and we have developed many of the techniques here in preparation to model such theories in the future. 
 
 We have not considered here the inclusion of fermionic bound states. One approach to include such states is the introduction of fermions in the bulk gravitational theory as dual states of baryons as has been outlined in \cite{Abt:2019tas,Erdmenger:2020flu}. That work we also leave for the future  but it would  lay the ground to investigate the top-partner spectrum of composite Higgs models.

\vspace{6pt}

\begin{appendix}
\section{Equations of Motions,  $N_f=2$  Split Masses}
\label[appendix]{appendix:Nf=2_no_str}
We use  $r_{i}^2=r_{i}^2(\rho)=L_i^2(\rho)+\rho^2$ ($i=d,u$) in the following. On calculating the masses, we took the limit of vanishing longitudinal $\phi_{v_i}$ and $\phi_{a_i}$. 

{\bf Scalars}

\begin{align}
    &\xi_{u,d}:\quad &\partial_\rho\left(\rho^3 \partial_\rho \xi_{u/d}(\rho)\right)+ \rho\left(-\Delta m_{u/d}^2-L_{u/d}(\rho ) \Delta m_{u/d}^{2\prime }+\frac{\rho ^2 M_{\xi_{u/d}}^2}{r_{u/d}^4}\right)\xi_{u/d}(\rho )&=0\\
    \nonumber
    &\xi_{1/2}:\quad &\partial_\rho\left(\rho^3 \partial_\rho \xi_{1/2}(\rho)\right)+\frac{\rho}{2}  \left[-\Delta m_{u}^2-\Delta m_{d}^2-L_u(\rho ) \Delta m_{u}^{2\prime}-L_d(\rho ) \Delta m_{d}^{2\prime}\right]\xi_{1/2}(\rho )&\\
   & &+\frac{1}{2}
   M_{\xi_{1/2}}^2 \rho ^3 \left(\frac{1}{r_d^4}+\frac{1}{r_{u}^4}\right) \left(\xi_{1/2}(\rho)\pm(\lu-\ld)\phi_{V_{2/1}}(\rho )\right)&=0\label{eq:Nf=2_TR_eoms_xi1}
\end{align}

{\bf Pseudo-Scalars}\\
\begin{align}
   & \pi_{u,d}:\quad &\partial_\rho\left(L_{u/d}^2(\rho)\rho^3\partial_\rho\pi_{u/d}(\rho)\right)+{L_{u/d}^2(\rho) M_{\pi_{u/d}}^2\rho^3\over r_{u/d}^4}\left(\pi_{u/d}(\rho)-\phi_{A_{u/d}}(\rho)\right) &=0\\   
   \label{eq:Nf=2_TR_eoms_pi1}
   \nonumber
    &\pi_{1/2}:\quad &\partial_\rho\left((\lu+\ld)^2\rho^3\pdr\pi_{1/2}(\rho)\right)+{M_{\pi_{1/2}}^2\rho^3\over 2}\left(\frac{1}{r_u^4}+\frac{1}{r_{d}^4}\right)(\lu+\ld)^2&\\
&&\times\left(\pi_{1/2}(\rho)-\phi_{A_{1/2}}(\rho)\right)&=0
\end{align}

{\bf Vectors}
\begin{alignat}{3}
  &V_{u,d}:\quad &\partial_\rho\left(\rho^3 \partial_\rho V_{u/d}(\rho)\right)+\frac{ \rho ^3 M_{V_{u/d}}^2}{ r_{u/d}^4} V_{u/d}(\rho )&=0\\
    \label{eq:Nf=2_TR_eoms_v1}
    &V_{1/2}:\quad &\partial_\rho\left(\rho^3 \partial_\rho V_{1/2}(\rho)\right)+\frac{\rho^3}{8}\left({1\over r_u^4}+{1\over r_d^4}\right)\left[4 M_{V_{1/2}}^2-g_5^2\left( L_u(\rho )-L_d(\rho )\right)^2 \right]V_{1/2}(\rho ) &=0
\end{alignat}

\begin{alignat}{3}
\label{eq:Nf=2_TR_eoms_vrho12}
     \nonumber
     & \phi_{v_{1/2}}:\quad &\partial_\rho\left(\rho^3 \partial_\rho\phi_{V_{1/2}}(\rho)\right)\pm\frac{\rho^3 g_5^2}{8}\left({1\over r_u^4}+{1\over r_d^4}\right)\left( L_u(\rho )-L_d(\rho )\right)&\\
     &&\times \lrb{\xi_{2/1}(\rho)\mp(\lu-\ld)\phi_{V_{1/2}}(\rho )} &=0\\
    &&4M_{\phi_{V_{1/2}}}^2\pdr\phi_{V_{1/2}}(\rho)\mp g_5^2(L_u(\rho )- L_d(\rho ))\pdr\xi_{2/1}(\rho)\pm g_5^2 \xi_{2/1}(\rho) \pdr(L_u(\rho )-L_d(\rho ))&=0
\end{alignat}

{\bf Axial-Vectors}
\begin{alignat}{3}
    &A_{u,d}:\quad &\partial_\rho\left(\rho^3 \partial_\rho A_{u/d}(\rho)\right)+ \frac{\rho ^3}{r_{u/d}^4}  \left(M_{A_{u/d}}^2-g_5^2 L_{u/d}(\rho )^2\right)A_{u/d}(\rho )&=0\\
    \label{eq:Nf=2_TR_eoms_arhoud}
  &\phi_{A_{u/d}}:\quad  &\pdbr{\phi_{A_{u/d}}(\rho)}+{\rho^3 g_5^2\over r_{u/d}^4}L_{u/d}(\rho)^2(\pi_{u/d}(\rho)-\phi_{A_{u/d}}(\rho))&=0\\
    && M_{\phi_{A_{u/d}}}^2\pdr\phi_{A_{u/d}}(\rho)-g_5^2L_{u/d}(\rho)^2\pdr\pi_{u/d}(\rho)&=0
  \end{alignat}

\begin{alignat}{3}
    &A_{1/2}:\quad &\partial_\rho\left(\rho^3 \partial_\rho A_{1/2}(\rho)\right)+\frac{\rho^3}{8}\left({1\over r_u^4}+{1\over r_d^4}\right)\left[ 4M_{A_{1/2}}^2-g_5^2\left( L_u(\rho )+L_d(\rho )\right)^2 \right]A_{1/2}(\rho ) &=0\\
    \label{eq:Nf=2_TR_eoms_arho12}
  &\phi_{A_{1/2}}:\quad  &\pdbr{\phi_{A_{1/2}}(\rho)}+{\rho^3 g_5^2\over 8}\invrud(L_u(\rho)+L_d(\rho))^2(\pi_{1/2}(\rho)-\phi_{A_{1/2}}(\rho))&=0\\
   && M_{\phi_{A_{u/d}}}^2\pdr\phi_{A_{1/2}}(\rho)-{g_5^2\over 4}(L_u(\rho)+L_d(\rho))^2\pdr\pi_{1/2}(\rho)&=0
\end{alignat}
From \cref{eq:Nf=2_TR_eoms_arhoud} and \cref{eq:Nf=2_TR_eoms_arho12} one recovers \cref{eq:Nf=2_TR_eoms_pi1}.
\section{Equations of Motions, $N_f=3$}
\label[appendix]{app:Nf=3}
As mentioned in \cref{Nf=3}, the equations of motion are sorted in groups $(u,1,2,3)$, $(4,5,6,7)$ and $s$. The groups $(u,1,2,3)$ and $s$ take the abelian form, but with $(L_u,r_u)$ and $(L_s,r_s)$ for $(u,1,2,3)$ and $s$ respectively. The group $(4,5,6,7)$ have mixed equations as the off-diagonal ones (1,2) in the two-flavour case, where (4,5) and (6,7) mixed in the same way as $N_f=2$ (1,2). We demonstrate this pattern with the scalars as an example.\\

The scalar fluctuations are
\begin{equation}
\label{eq:Nf=3_scalar_u123}
     \left[\partial_\rho\left(\rho^3\partial_\rho\right)-\Delta m_u^2(\rho)\rho-L_u(\rho)\frac{\partial \Delta m_u^2(\rho)}{\partial L_u}\rho+\rho^3\frac{M_{\sigma_i}^2}{r_u^2}\right]\sigma_i(\rho)=0,\quad i=u,1,...,3
 \end{equation}
\begin{equation}
    \left[\partial_\rho\left(\rho^3\partial_\rho\right)-\Delta m^2_s(\rho)\rho-L_s(\rho)\frac{\partial \Delta m^2_s(\rho)}{\partial L_s}\rho+\rho^3\frac{M_{\sigma_i}^2}{r_s^2}\right]\sigma_i(\rho)=0,\quad i=s
\end{equation}
where $\sigma_u$ and $\sigma_s$ are defined in \cref{eq:N_f=3_redef_fields}. To get the equations for off-diagonal ones (4,5,6,7), one can simply take \cref{eq:Nf=2_TR_eoms_xi1} and change the pairs
\begin{equation}
    \begin{aligned}
   & (1,2)\rightarrow (4,5),\, (6,7)\\
   & (L_u,L_d)\rightarrow(L_u,L_s).
    \end{aligned}
\end{equation}
The other fluctuations' equations follow the same logic.
\acknowledgments{In completing this paper, Y.L was supported by a fellowship of the `Studienstiftung des deutschen Volkes'. N.E.'s work was supported by the STFC consolidated grant  ST/T000775/1. The authors would like to thank A. Karch and K. Landsteiner for discussions.}
\end{appendix}

\bibliographystyle{utphys}
\bibliography{main}

\end{document}